\documentclass[a4paper,11pt]{article}
\pdfoutput=1 

\usepackage{jcappub} 

\usepackage[T1]{fontenc} 
\usepackage{color}
\usepackage{mathtools}
\usepackage{aas_macros}
\usepackage[normalem]{ulem}

\newcommand{\avg}[1]{\ensuremath{\langle #1 \rangle}}
\newcommand{\bma}{\begin{math}}
\newcommand{\ema}{\end{math}}
\newcommand{\beq}{\begin{equation}}
\newcommand{\eeq}{\end{equation}}
\newcommand{\beqa}{\begin{eqnarray}}
\newcommand{\eeqa}{\end{eqnarray}}
\newcommand{\bc}{\begin{center}}
\newcommand{\ec}{\end{center}} 
\newcommand{\bit}{\begin{itemize}}
\newcommand{\eit}{\end{itemize}}

\usepackage{amsmath}
\usepackage[utf8]{inputenc}

\usepackage{hyperref,breakurl}
\hypersetup{colorlinks=true, citecolor=blue, urlcolor=blue, linkcolor=blue}
\usepackage[usenames,dvipsnames]{xcolor}
\usepackage{dcolumn}
\usepackage{bm}
\usepackage{array}
\usepackage{graphicx}
\usepackage{graphics}
\usepackage{slashed}
\usepackage{subfig}

\title{Probing Cosmic Magnetism with Rotation Measure-Squared-Galaxy Cross-Correlations}

\author[a, b]{Zekai Zhang}
\author[a]{and Adam Lidz}

\affiliation[a]{Department of Physics and Astronomy, University of Pennsylvania,\\Philadelphia, PA 19104, USA}
\affiliation[b]{Department of Mechanical Engineering and Applied Mechanics, University of Pennsylvania,\\Philadelphia, PA 19104, USA}

\emailAdd{zzekai@sas.upenn.edu}
\emailAdd{alidz@sas.upenn.edu}

\date{\today}

\abstract{
We present a new approach for extracting information about cosmic magnetic fields using cross-correlations between extragalactic Faraday rotation measure (RM) catalogs and galaxy surveys. Specifically, we propose measuring the two-point cross-correlation between RM squared, ${\rm RM}^2$, towards background sources and the projected density field of foreground galaxies, $\langle {\rm RM}^2 \times {\rm g} \rangle$, as a function of transverse separation. This statistic is analogous to the ``projected fields''  estimator used for the kinetic Sunyaev-Zel'dovich (kSZ) effect, $\langle {\rm kSZ}^2 \times {\rm g} \rangle$. Our estimator avoids contamination, on average, from RM contributions originating in the background source or the Milky Way, and is also free from the noise bias that arises when correlating the absolute value of the RMs with galaxies. Moreover, by binning in foreground galaxy redshifts, $\langle {\rm RM}^2 \times {\rm g} \rangle$ enables a tomographic reconstruction of the redshift evolution of large-scale cosmic magnetic fields. We model this statistic using the \textsc{Illustris-TNG} cosmological magnetohydrodynamic simulations and compare with approximate analytic predictions. In analogy with the kSZ case, we show that $\langle {\rm RM}^2 \times {\rm g} \rangle$ can be related to a bispectrum involving two copies of the electron-density--weighted magnetic field strength and one of the galaxy overdensity. On large scales, the correlation approximately traces the shape of the electron-galaxy two-point function, with an amplitude governed by the average electron-weighted magnetic field strength and the smoothing scale of the RM measurements. In \textsc{Illustris-TNG}, the effective field strength is primarily set by the magnetic field amplitudes within the inner regions of galaxy-hosting dark matter halos. It increases towards low redshift, driven by dynamo amplification and magnetized outflows.  
Our forecasts suggest that $\langle {\rm RM}^2 \times {\rm g} \rangle$ is detectable at high significance with current galaxy surveys and future RM catalogs from the SKA, offering a tomographic probe of large-scale magnetic fields across cosmic time. }


\begin{document}
\maketitle
\flushbottom

\section{Introduction}
\label{sec:intro}

Magnetic fields are found throughout the universe, in planets, stars, compact objects, galaxies, clusters of galaxies, and in the filaments of the cosmic web \citep{Subramanian:2015lua,beck_galactic_2009,widrow_origin_2002}. Indirect evidence suggests that magnetic fields may even pervade much of the low-density intergalactic medium \citep{Taylor:2011bn}. Despite their ubiquity, the origin, evolution, and spatial structure of cosmic magnetic fields remain poorly understood. 
Ideally, we would map the ``magnetic cosmic web'' in three dimensions, tracing its structure alongside the sheets, filaments, halos, and voids of the large-scale structure distribution. 
A better observational census of the magnetic cosmic web can help in addressing some of the major open questions in cosmic magnetism \citep{widrow_origin_2002,beck_galactic_2009, Subramanian:2015lua}. Which processes seed the universe with magnetic fields? Do magnetic fields form before galaxies and provide an additional fossil relic of the early universe? How are seed magnetic fields amplified? How do magnetic fields impact the formation of stars, galaxies, and cosmic structure? What role do outflows from galaxies and active galactic nuclei (AGN) play in spreading magnetic fields throughout the universe?

Fortunately, there are outstanding prospects for improving our census of cosmic magnetic fields in the near future and making progress on these open questions. A key observational probe is Faraday rotation: the polarization angle of a linearly polarized source rotates by an angle proportional to wavelength-squared and to the so-called rotation measure (RM) \citep{Rybicki86}. The RM is obtained by integrating, along the line of sight to the background source, the product of the electron density and the magnetic field component parallel to the sightline. Excitingly, the number of extragalactic RM measurements is growing rapidly \citep{haarlem_lofar_2013,osullivan_faraday_2023}, and the Square Kilometer Array (SKA) is expected to provide $\sim 10^7$ extragalactic RM measurements, spanning nearly the full sky \citep{gaensler_origin_2004,gaensler_cosmic_2008}.

One issue with RM measurements, however, is that they provide only an integral constraint. Thus, each RM generally receives contributions from magnetized plasma within the source, in the Milky Way, as well as from other intervening material. The latter contribution depends on the particular path traced through the large-scale structure, with some sightlines passing through halos and filaments, while others traverse mainly void regions. 

An elegant approach for sharpening the interpretation of RM measurements is to pursue cross-correlations with tracers of large-scale structure, such as galaxy surveys \citep{lee_detection_2010,stasyszyn_measuring_2010,amaral_constraints_2021, Kolatt:1997xu}. 
A key advantage of this technique is that only RM contributions from near the redshifts of the galaxies surveyed lead to an average cross-correlation, while Milky Way and intrinsic source RM contributions do not produce an average correlation signal. However, a complication is that direct RM-galaxy correlations cancel because the RM field can be either positive or negative. 
Previous work hence considered cross-correlations between the absolute value of the RM field and the galaxy distribution. 
Drawbacks of this approach include that $|\mathrm{RM}|$ is not differentiable at $\mathrm{RM}=0$, and that it is difficult to relate this cross-correlation to the underlying power spectra and bispectra of the magnetic field, electron density, and galaxy distributions. Furthermore, this statistic is biased by noise, as we will discuss.

We propose here, for the first time, a promising alternative: the cross-correlation function between the RM-squared field and the galaxy distribution, measured as a function of the transverse separation between the foreground galaxy distribution and background RM$^2$ measurements. 
 We will often denote this by the shorthand $\langle \mathrm{RM}^2 \times \mathrm{g} \rangle$ in what follows. This statistic avoids cancellations and noise bias, and is analogous to the ``projected fields'' estimator employed successfully in the context of the kinetic Sunyaev-Zel'dovich effect, $\langle \mathrm{kSZ}^2 \times \mathrm{g} \rangle$ \citep{Dore:2003ex, Hill:2016dta, Ferraro:2016ymw}. 
 
Next, we simulate this statistic. Along with observational advances in extragalactic RM catalogs, there has also been important progress in cosmological magnetohydrodynamic simulations \citep{Dolag:2008ya, pakmor_magnetohydrodynamics_2011, Vazza14, pillepich_first_2018, marinacci_first_2018, Garaldi21}. Notably, the publicly available \textsc{Illustris-TNG} simulations provide realizations of the magnetized cosmic web across cosmological volumes, along with models for the distributions of baryons, galaxies, and the underlying dark matter \citep{Weinberger17,Pillepich18,nelson_first_2019,marinacci_first_2018}. This allows us to explore the information content of the $\langle \mathrm{RM}^2 \times \mathrm{g} \rangle$ cross-correlation function, and to sharpen our understanding of the science that can be extracted from future measurements. We also develop an analytic framework for interpreting the simulation results and those of future observations. 

Finally, we forecast the prospects for measuring $\langle \mathrm{RM}^2 \times \mathrm{g} \rangle$ correlations using current and upcoming extragalactic RM catalogs and galaxy survey data. These calculations help quantify the expected science return of our proposed analyses and may provide useful guidance for optimizing future survey strategies. 

Section~\ref{sec:method} describes the $\langle \mathrm{RM}^2 \times \mathrm{g} \rangle$ statistic used in this work. There we also present our methodology for constructing simulated RM maps and for estimating cross-correlation statistics from the simulations. In Section~\ref{sec:results} we show calculations of our
$\langle \mathrm{RM}^2 \times \mathrm{g} \rangle$ statistic from the \textsc{Illustris-TNG} simulations, exploring the signal's scale and redshift dependence. We also compare with the approximate analytic model, mentioned earlier, which can explain some of the main features of the simulation results. 
In addition, we investigate the scale and redshift dependence of the $\langle |\mathrm{RM}| \times \mathrm{g} \rangle$ statistic in \textsc{Illustris-TNG}. Next, Section~\ref{sec:uncertainties} forecasts the $\langle \mathrm{RM}^2 \times \mathrm{g} \rangle$ detection prospects for current and forthcoming RM catalogs and galaxy surveys. We summarize our conclusions in Section~\ref{sec:conclusion}. Appendix~\ref{app:a} provides some details regarding the 
$\langle \mathrm{RM}^2 \times \mathrm{g} \rangle$ statistic and its relation to the underlying statistical properties of the cosmic magnetic fields, the electron density distribution, and the galaxy fluctuation field. Appendix~\ref{app:b} justifies an approximate expression used in this work to estimate error bars for future $\langle \mathrm{RM}^2 \times \mathrm{g} \rangle$ measurements. Finally, Appendix~\ref{app:c} tests how well the simulation results converge with increasing resolution.

\section{Methods} \label{sec:method}

First, we introduce the RM$^2$-galaxy cross-correlation statistic, and given an overview of its properties. We then describe our approaches for constructing RM maps and estimating $\langle \mathrm{RM}^2 \times \mathrm{g} \rangle$ from the \textsc{Illustris-TNG} simulations. We also compare the RM$^2$-based estimator with the absolute-value alternative, $\langle |\mathrm{RM}| \times \mathrm{g} \rangle$, employed in earlier work. 

\subsection{Simulating Rotation Measure Maps}

Our methodology starts by constructing an RM map. That is, we  
deposit discrete RM measurements towards background sources onto a pixelized grid, and smooth the map before squaring the field. The resulting map is
\begin{equation}
\rm{RM}(\boldsymbol{\theta}) = \sum_i F(\boldsymbol{\theta} - \boldsymbol{\theta_i}) \rm{RM}_i(\boldsymbol{\theta_i}),
\end{equation}
where $\rm{RM}_i(\boldsymbol{\theta_i})$ is the RM measurement towards a discrete background source in direction $\boldsymbol{\theta_i}$, and $F$ is a filter describing the combined effects of pixelization and smoothing.

{This map-based approach differs from a discrete estimator in which
$\rm{RM}^2_i$ values are directly correlated with galaxy positions. 
 The discrete estimator probes small-scale structures along the line-of-sight, with transverse scales set by the larger of the RM source size and the coherence length of the electron-density weighted magnetic field. In contrast, the map-based approach introduces an explicit smoothing scale, yielding a well-defined transverse averaging scale which can be consistently matched between simulations and observations. 
 For brevity, we leave the smoothing scale dependence implicit in most of what follows, but discuss it quantitatively in Section \ref{sec:fiducial_rm}.

The Faraday RM along a line of sight (LoS) in the sky direction $\boldsymbol{\theta}$ is proportional to an integral of the electron–density–weighted line-of-sight magnetic field component:
\begin{equation}
\label{eq:rm_integral}
\mathrm{RM}(\boldsymbol{\theta}) = 8.12 \times 10^5~\mathrm{rad \, m^{-2}} \int_0^{z_s} dz   \frac{dl}{dz}   \frac{n_e(\boldsymbol{\theta},z) B_\parallel(\boldsymbol{\theta},z)}{(1+z)^2},
\end{equation}
where $z_s$ is the redshift of the source. 

In general, a given map pixel will contain sources spanning a range of redshifts. However, since we focus on cross-correlations with lower-redshift galaxy samples, the precise source redshift distribution is of limited importance as long as the sources are at larger redshift than the galaxies in the cross-correlation samples. 

Here, $n_e$ is the proper electron number density in cm$^{-3}$, $B_\parallel$ is the proper magnetic field component along the LoS in $\mu$G, and $dl/dz$ is the proper distance per unit redshift in Mpc units. The $(1+z)^{-2}$ factor arises
because of the $\lambda^2$ dependence of Faraday rotation, and accounts for the wavelength stretching between the plasma frame and the frame of an observer on Earth. 

It will be convenient to work primarily in comoving coordinates. Let us express the proper electron abundance as $n_e= \avg{n_e(z=0)}(1+z)^3[1+\delta_e]$, where $\delta_e$ describes the fractional electron abundance fluctuations, and $\avg{n_e(z=0)}$ is the mean electron number density at $z=0$.
After moving to an integral over comoving length scales, the redshift scaling factors in the RM integral cancel, and we have:
\begin{equation}
\label{eq:rm_chi}
\mathrm{RM}(\boldsymbol{\theta}) = W_{\mathrm{RM}} \int d\chi \Theta(\chi_s-\chi) [1 + \delta_e(\chi \boldsymbol{\theta}, \chi)] B_\parallel(\chi \boldsymbol{\theta}, \chi), 
\end{equation} 
 where $W_{\mathrm{RM}} = 8.12 \times 10^5 \langle n_e(z=0) \rangle~\mathrm{rad \, m^{-2} \mu G^{-1} Mpc^{-1}}$ is a constant pre-factor. The top-hat function, $\Theta(\chi)$, simply enforces the upper cut-off in the integration at the comoving distance to the source, $\chi_s$.
 In the flat-sky approximation, adopted throughout this work, angular positions and transverse comoving separations are related by $r_\perp = \chi(z) \boldsymbol{\theta}$.

In our analysis, we construct RM maps in each coeval (fixed redshift) simulation snapshot by evaluating Eq.~\ref{eq:rm_chi}. 
As mentioned earlier, $\mathrm{RM}$-galaxy cross-correlations would suffer from cancellations since the $\mathrm{RM}$ field can be either positive or negative. Here, we circumvent this issue by working with the squared-RM field: 
\begin{equation}
A(\boldsymbol{\theta}) \equiv \mathrm{RM}^2(\boldsymbol{\theta}).
\end{equation}
We then cross-correlate this with the projected galaxy overdensity field $\delta_g(\boldsymbol{\theta})$ in different galaxy redshift bins. 

\subsection{The $\mathrm{RM}^2$-galaxy cross-correlation statistic and its properties}

The $\langle \mathrm{RM}^2 \times \mathrm{g} \rangle$ cross-correlation function can be related to a bispectrum involving the underlying electron-density, magnetic field, and galaxy distribution fluctuations. As alluded to the Introduction, the relationship can be derived in close analogy to the case of the $\langle \mathrm{kSZ}^2 \times \mathrm{g} \rangle$ signal \citep{Dore:2003ex}. We leave the details of the derivation to Appendix \ref{app:a}, but summarize key points here. 

It is helpful to turn to harmonic space, assuming the flat-sky and Limber approximations throughout. The Fourier transform of $A(\boldsymbol{\theta})$ involves the convolution between the transforms of the RM fields:
\begin{equation}
\label{eq:a_conv}
A(\boldsymbol{\ell}) = \int \frac{d^2\ell_1}{(2\pi)^2}   \mathrm{RM}(\boldsymbol{\ell} - \boldsymbol{\ell}_1)   \mathrm{RM}(\boldsymbol{\ell}_1).
\end{equation}
Note that here we neglect any smoothing of the RM field from, for example, the finite angular resolution of the RM observations. We investigate the impact of RM-smoothing in Section \ref{sec:fiducial_rm}.
The projected galaxy fluctuation field can be described by
\begin{equation}
\label{eq:proj_gal}
\delta_g(\boldsymbol{\theta}) = \int d\chi   W_g(\chi) \delta_g(\chi \boldsymbol{\theta}, \chi),
\end{equation}
where $W_g(\chi)$ gives the probability per unit comoving length to find a galaxy at comoving distance $\chi$.

As shown in Appendix \ref{app:a}, the angular cross-power spectrum of the $A(\boldsymbol{\theta})$ and $\delta_g(\boldsymbol{\theta})$ fields is given by:
\begin{equation}
\label{eq:ca_g_triangle}
C_{A,g}(\ell) = W_{\mathrm{RM}}^2 \int \frac{d\chi}{\chi^4}   W_g(\chi)   \mathcal{T}\left(k = \frac{\ell}{\chi}; z\right),
\end{equation}
where the triangle power spectrum $\mathcal{T}(k)$ is defined by an integral over a bispectrum:
\begin{equation}
\label{eq:triangle_power}
\mathcal{T}(k; z) = \int \frac{d^2q}{(2\pi)^2}   \mathcal{B}_{\tilde{B}_{\perp,\parallel}; \tilde{B}_{\perp,\parallel}; g}(\boldsymbol{k - q},  \boldsymbol{q}, -\boldsymbol{k}; z).
\end{equation}
This expression has the same general form as an analogous result in the case of the kSZ projected fields signal \citep{Dore:2003ex, Ferraro:2016ymw}. Here $\tilde{B}_\parallel(\boldsymbol{x}) \equiv [1 + \delta_e(\boldsymbol{x})] B_\parallel(\boldsymbol{x})$ is the LoS component of the electron-density--weighted magnetic field. 
The $\perp$ subscript refers to the projection perpendicular to the wavevector $\boldsymbol{k}$, so that $\tilde{B}_{\perp,\parallel}(\boldsymbol{k})$ denotes the component of $\boldsymbol{\tilde{B}}$ that is perpendicular to $\boldsymbol{k}$ but parallel to the LoS. 
The 3D bispectrum $\mathcal{B}$ describes correlations between two copies of the electron-density--weighted magnetic field, $\tilde{B}_{\perp,\parallel}$, and the galaxy distribution $\delta_g$.   

The real-space angular correlation function is the Fourier transform of $C_{A,g}(\ell)$:
\begin{equation}
\label{eq:w_ag_real}
w_{\mathrm{RM^2,g}}(\theta) = \int \frac{d\ell}{2\pi}   \ell   J_0(\ell\theta)   C_{A,g}(\ell),
\end{equation}
where our notation has moved back to $\mathrm{RM^2}$ from ``$A$'', as $A$ is mainly convenenient in harmonic space. 
Using Eq.~\ref{eq:ca_g_triangle}, this can be written as:
\begin{equation}
\label{eq:wag_triangle}
w_{\mathrm{RM^2,g}}(\theta) = W_{\mathrm{RM}}^2 \int d\chi \Theta(\chi_s-\chi)  \frac{W_g(\chi)}{\chi^2} \int \frac{dk}{2\pi}   k   J_0(k \chi \theta)   \mathcal{T}(k; z).
\end{equation}
Alternatively, the same correlation can be expressed as a line-of-sight projection of the 3D correlation between the squared $\tilde{B}_\parallel$ field and the galaxy overdensity:
\begin{equation}
\label{eq:wag_coev}
w_{\mathrm{RM^2,g}}(\theta) = W_{\mathrm{RM}}^2 \int d\chi \Theta(\chi_s-\chi)  W_g(\chi) 
\int_{-\infty}^{\infty} dx_\parallel   \xi_{\tilde{B}_\parallel^2, g} \left(\sqrt{x_\parallel^2 + \chi^2 \theta^2}; z\right).
\end{equation}
Following Eq.~\ref{eq:wag_coev}, we define a coeval correlation function:
\begin{equation}
\label{eq:wag_weight}
w_{\mathrm{RM^2,g}}(\theta) = \int d\chi \Theta(\chi_s-\chi) W_g(\chi)   w_{\mathrm{coeval}}(\chi \theta; z),
\end{equation}
where
\begin{equation}
\label{eq:wcoeval}
w_{\mathrm{coeval}}(r_\perp; z) \equiv W_{\mathrm{RM}}^2 \int_{-\infty}^{\infty} dx_\parallel   \xi_{\tilde{B}_\parallel^2, g} \left(\sqrt{x_\parallel^2 + r_\perp^2}; z\right),
\end{equation}
and $r_\perp = \chi(z) \theta$. 
In the limit of a narrow galaxy redshift bin, so that $W_g(\chi)$ is sharply peaked at a single redshift, $z < z_s$, measurements of $\langle \mathrm{RM}^2 \times g \rangle$ directly probe the coeval projected correlation function of Eq.~\ref{eq:wcoeval}. In what follows, we generally refer to this simply as $w_{\mathrm{RM^2,g}}(r_\perp;z)$. The projected correlation function above is related to the triangle power spectrum through Eq.~\ref{eq:wag_triangle}, which itself is an integral over a bispectrum (Eq.~\ref{eq:triangle_power}). 
Eq.~\ref{eq:wag_weight} also shows that the angular cross-correlation function for an arbitrary foreground galaxy redshift distribution can be obtained  by taking an appropriate weighted average of the coeval projected correlation functions at different redshifts. This result assumes the Limber approximation. In practice, this means that the full angular correlation function, for an arbitrary galaxy redshift distribution,  can be modeled by combining coeval projected correlation functions measured from a sequence of simulation snapshots.

\subsection{Mock RM Maps and Statistics from \textsc{Illustris-TNG}}
\label{sec:illustris_rm}

We use cosmological magnetohydrodynamical (MHD) simulations from the \textsc{Illustris-TNG} project\footnote{\url{http://www.tng-project.org}} to construct mock RM maps and to model their cross-correlations with the galaxy distribution. The \textsc{Illustris-TNG} simulation suite are state-of-the-art simulations of galaxy formation in its proper cosmological context: these simulations self-consistently follow the coupled evolution of dark matter, gas, stars, active galactic nuclei (AGN), and magnetic fields across cosmic time \citep{nelson_first_2018, springel_first_2018, pillepich_first_2018, naiman_first_2018, marinacci_first_2018}. The simulation suite includes three different boxsizes: TNG50, TNG100, and TNG300, with respective comoving side lengths of 50, 100, and 300 Mpc, each of which has been run at three different resolution levels. For example, TNG300-1 represents the highest-resolution version of the TNG300 simulation, with $2500^3$ Voronoi gas cells, while TNG300-3 is the lowest-resolution case, containing 64 times fewer resolution elements.

In most of this work, we adopt the TNG300-3 simulation. The large volume of TNG-300 allows us to estimate the correlation functions of interest on fairly large scales. Although this run has lower resolution than the TNG300-1 and TNG300-2 simulations, it allows faster estimates of the cross-correlation functions. In Appendix \ref{app:c} we test the impact of resolution on our results for select cases, and find that the lower resolution runs still provide fairly accurate results. 

For each simulation snapshot from $z \sim 0-4$, we construct a map of RM measurements on a pixelized grid with co-moving pixels of $0.41$ Mpc/$h$ side length. To construct this map, we first interpolate the electron abundance and magnetic fields from the simulated fluid elements onto a regular 3D Cartesian grid. The chosen mesh scale of $0.41$ Mpc/$h$ is directly comparable to the expected mean SKA RM source separation at a representative redshift of $z \sim 0.2$. We perform a similar interpolation for the dark matter particle and subhalo positions. During the construction of the Cartesian grids, all data fields are converted into physical units. Electron density is expressed in $\mathrm{cm^{-3}}$, while the magnetic field unit is $\mu G$. 

Next, we use Eq.~\ref{eq:rm_chi} to estimate the coeval RM field across each snapshot. The RM field is estimated for each of $500^2$ evenly spaced lines of sight per snapshot, with each LoS pointing along the $z$-axis, normal to the $x-y$ plane. We also perform similar LoS-projections for the electron and halo overdensity fields, and the $\boldsymbol{\hat{z}}$-component of the magnetic field. 

Figure~\ref{fig:data} shows a visualization of the resulting projected halo and electron overdensity distributions, along with the magnetic field strength, and the coeval RM field. Specifically, the magnetic field strength is characterized here by the absolute value of the $\boldsymbol{\hat{z}}$-component of the magnetic field, $|B_z|$, after projection.
All four fields broadly trace the same filamentary large-scale structure, as expected. By eye, correlations can easily be discerned between the simulated RM$^2$ field (bottom right) and the projected halo distribution (upper left).  
The qualitative behavior seen here partly motivates our RM$^2$-galaxy cross-correlation analysis, which may help to infer the statistical properties of these fields. Another notable feature of the maps is the large contrast between the regions with large RM$^2$ (and hence large $|B_z|$), and those with low RM$^2$ (and hence small $|B_z|$). The distributions of the projected magnetic field strengths and the RMs are broad and non-Gaussian. Further, one can see that the regions of enhanced $|B_z|$ and RM$^2$
are more spatially-extended than the corresponding electron overdensities. This reflects the amplification of magnetic fields in collapsed regions and the outflow-driven transport of magnetized plasma into the surrounding circumgalactic and intergalactic media \citep{aramburo-garcia_magnetization_2021}. 

\begin{figure*}[ht!]
    \centering 
    \includegraphics[width=0.49\linewidth]{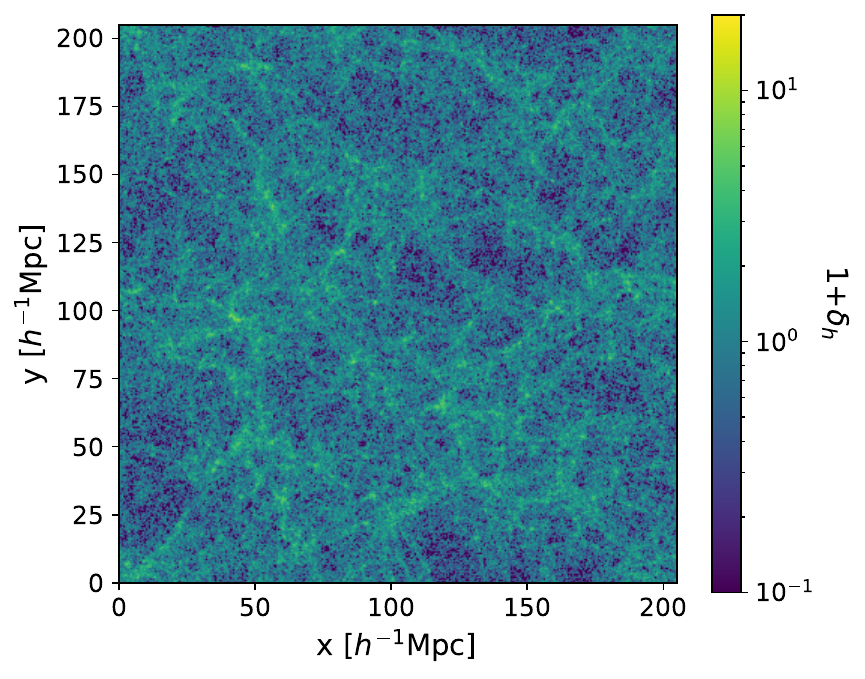}  
    \includegraphics[width=0.48\linewidth]{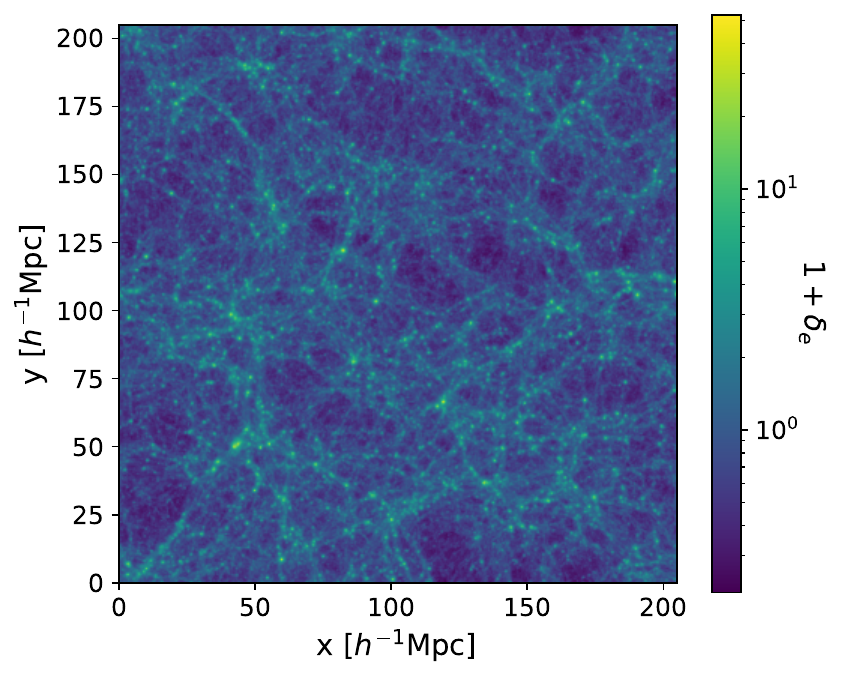}
    \includegraphics[width=0.49\linewidth]{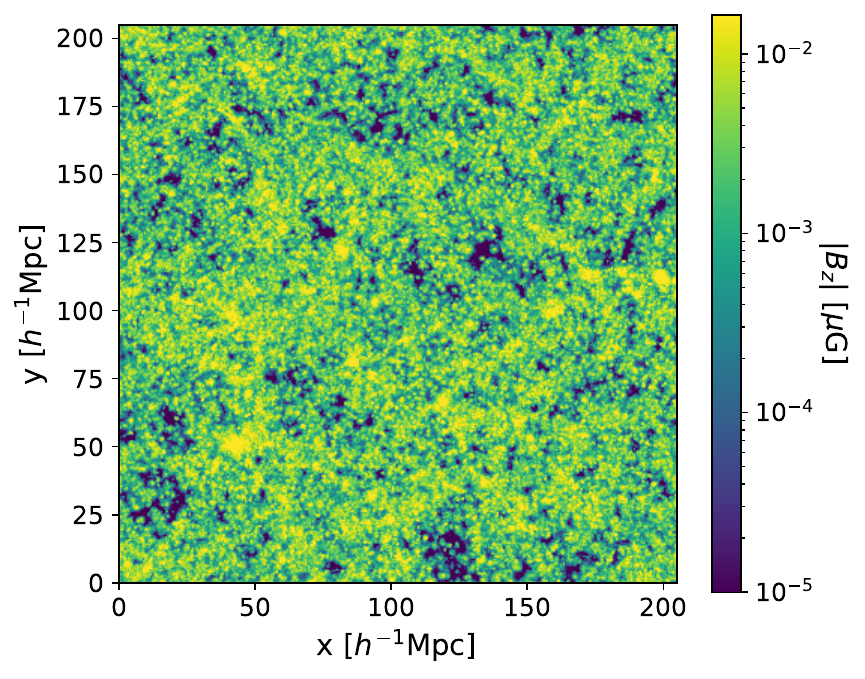}
    \includegraphics[width=0.49\linewidth]{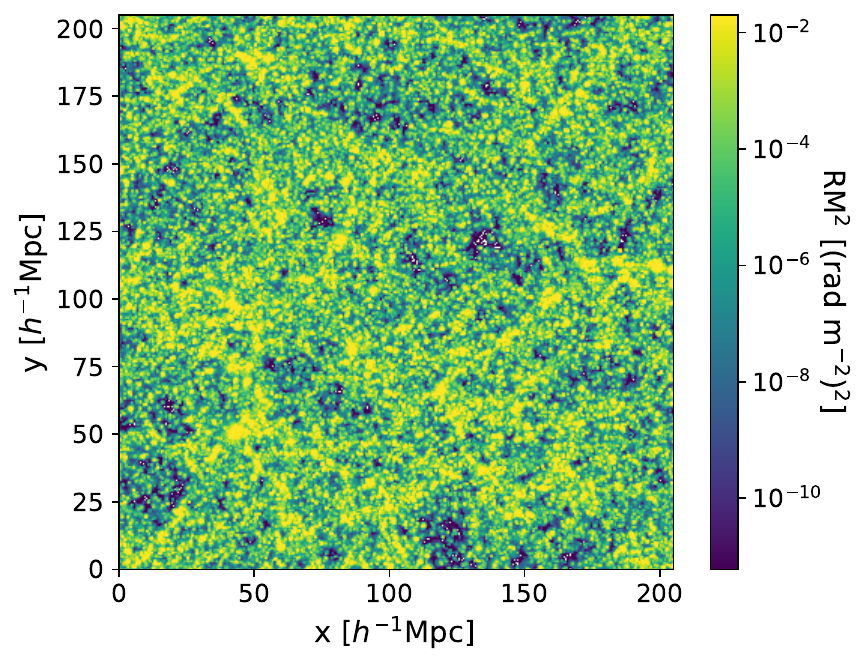}
    \caption{Projected fields in TNG300-3 at $z=0$. In each panel, the fields are projected along the full length of the simulation box, $L_{\rm box} = 205$ $h^{-1}$ Mpc, and the slice widths match $L_{\rm box}$.   
    \textit{Top left}: the projected halo overdensity field. \textit{Top right}: the electron overdensity field. \textit{Bottom left}: the absolute value of the projected line-of-sight component of the magnetic field. \textit{Bottom right}: the RM$^2$ contributions from magnetized plasma across the coeval box. The projected maps illustrate how the large-scale dark matter halo distribution, the electron density, magnetic field strength, and RM$^2$ fields all trace the same underlying distribution of large-scale structure. This motivates measuring cross-correlations between the halo (or galaxy) distribution and RM$^2$ measurements to infer some of the statistical properties of these fields. The ranges on the RM$^2$ and magnetic field strength color bars encapsulate 5\%-95\% of the simulation pixels.     }
    \label{fig:data}
\end{figure*}

\begin{figure*}[ht!]
    \centering 
    \includegraphics[width=0.32\linewidth]{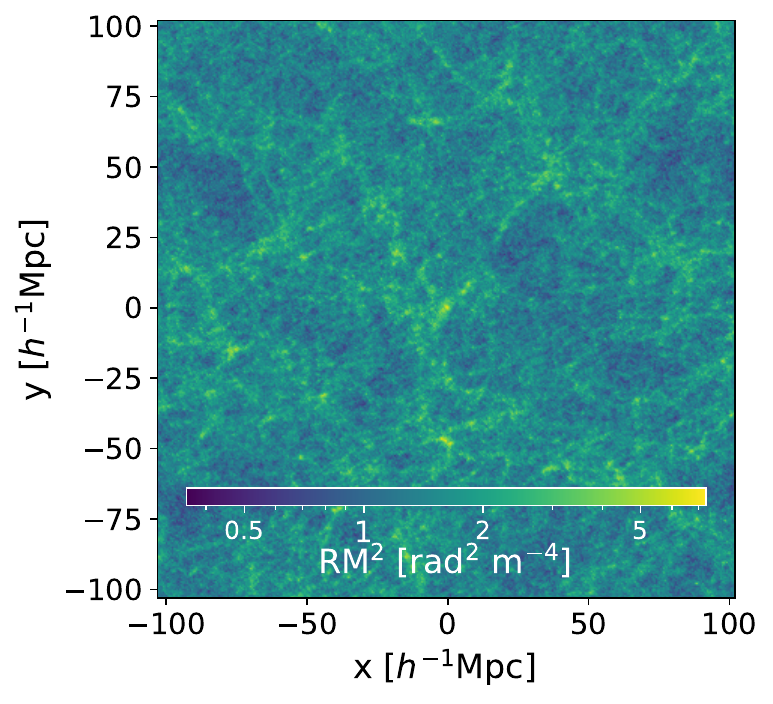}
    \includegraphics[width=0.32\linewidth]{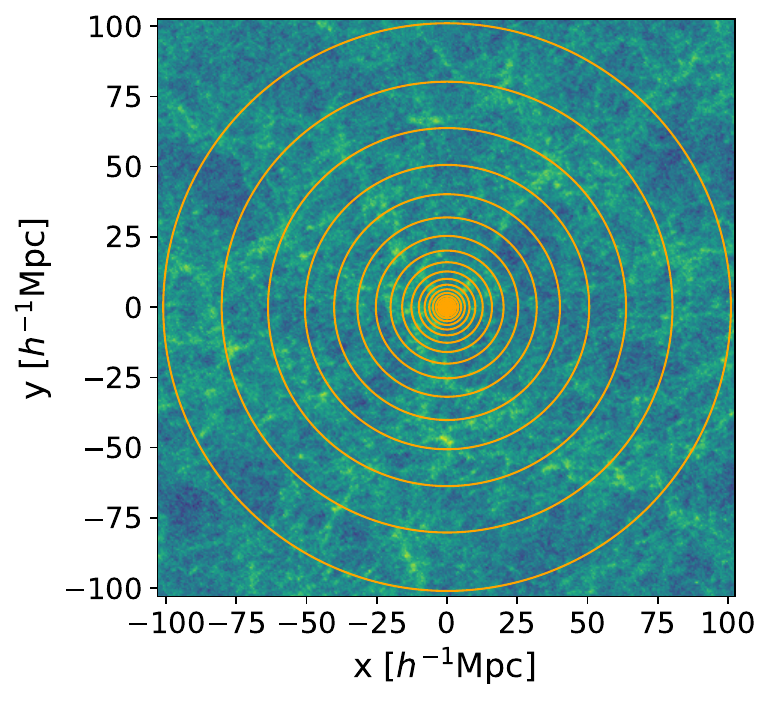}
    \includegraphics[width=0.32\linewidth]{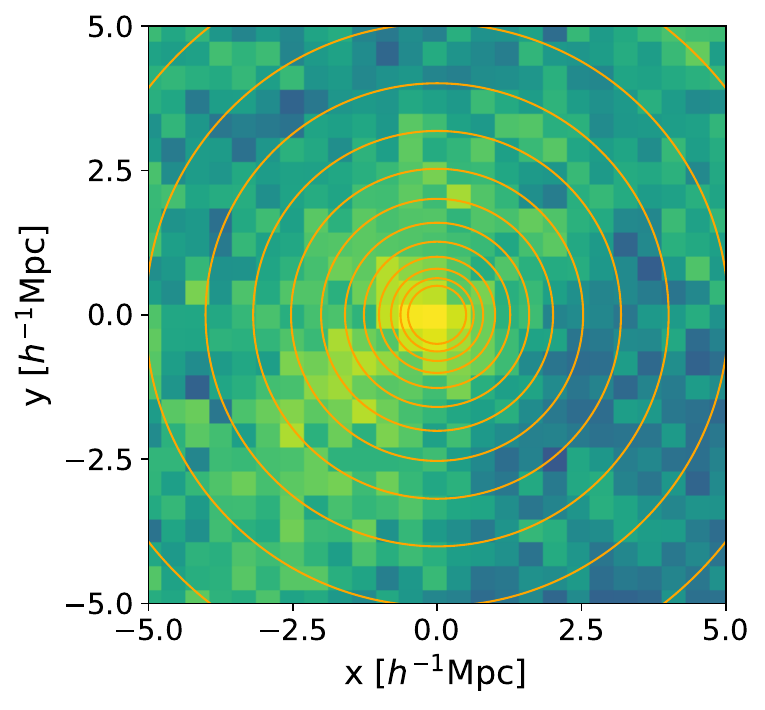}
    \caption{Schematic illustration of the stacking method and radial binning used in computing the coeval cross-correlation functions. 
    The radial bins increase logarithmically from the halo center out to half of the box size (102.5 Mpc/$h$). \textit{Left}: The RM$^2$ grid stacked around 391,144 halos from TNG300-3 at $z = 0$. \textit{Middle}: The radial binning employed, showing rings from $r$ to $r+dr$. \textit{Right}: Zoomed-in view of the stacked grid and inner radial bins near the stacking center. The excess $\rm{RM}^2$ around the stacked simulated halos is visible by eye. Measuring this signal in upcoming data sets as a function of scale, redshift, and potentially galaxy properties should help extract a wealth of information about cosmic magnetic fields. 
    } 
    \label{fig:stack}
\end{figure*}

We then turn to estimate the coeval cross-correlation signals across each available TNG300-3 snapshot from $z = 0-4$. In most of this work, we study simulated RM$^2$-halo cross-correlations, rather than cross-correlations with simulated galaxies. In this study, we confine our attention to scales larger than the size of individual galaxies. In this large-scale regime, we expect the galaxies and dark matter halos to trace the same large-scale structure, and so RM$^2$-halo cross-correlations will differ from RM$^2$-galaxy cross-correlations mainly in overall amplitude. That is, we expect a linear-biasing description to be a reasonable approximation on most scales and redshifts of interest. In this regime, using the full simulated halo catalog provides a less-noisy estimate of the simulated cross-correlation signals than is possible from the less abundant simulated galaxy populations.  Generally, we expect the RM$^2$-halo cross-correlations to be lower in amplitude at roughly the factor of $\sim 2$ level than RM$^2$-galaxy correlations, with the precise factor depending on the galaxy sample and redshift. The lower amplitude here reflects the smaller clustering bias around the low-mass halos included in the \textsc{Illustris-TNG} catalogs. Where relevant, we provide further comparisons between the expected RM$^2$-halo and RM$^2$-galaxy cross-correlation signals in what follows.

In order to estimate $w_{\mathrm{RM}^2, \mathrm{g}}(r_\perp;z)$ from the simulations, we employ a stacking-based method. Specifically, for each simulation snapshot, we shift the gridded RM field such that each halo's position is centered, using the simulation's periodic boundary conditions. We stack the shifted RM map around the central halo for all of the halos in the simulation box.
We then determine the average value of the halo-centered RM$^2$ field as a function of transverse separation, computing averages by summing over all simulation pixels in each bin and over all of the halos in the snapshot of interest. Finally, we subtract off the global average RM$^2$ value to determine the excess RM$^2$ around the simulated halos.\footnote{Note that in the simulations we only consider the coeval RM$^2$ produced by magnetized plasma at the redshifts of the simulated halos. In the case of actual data, the RM$^2$ fields will contain contributions from magnetized plasma at a range of redshifts, generally beyond those spanned by the galaxies in the survey being used for cross-correlation measurements.  One can still determine the excess RM$^2$ around the galaxies by subtracting estimates of RM$^2$ around {\em random} locations.} This stacking method is equivalent to a direct estimate of the average of the product of RM$^2$ and $\delta_{\mathrm{g}}$ over pixel pairs, as a function of their transverse separation.    

Figure~\ref{fig:stack} shows a schematic illustration of our $w_{\mathrm{RM}^2, \mathrm{g}}(r_\perp;z)$ estimation procedure. The excess $\mathrm{RM}^2$ around each halo center is clearly visible, with the rings indicating the radial bins used in the analysis. The stacked maps also exhibit noticeable azimuthal asymmetries. These arise primarily because the RM$^2$ field receives dominant contributions from rare, highly magnetized halos. It is also partly due to the finite boxsize of \textsc{Illustris-TNG}. The simulation spans a smaller effective sky area than surveys such as the SKA, which cover much larger sky fractions. As a result, stacked measurements from real data are expected to appear more azimuthally symmetric (up to measurement noise).

As a test of our correlation-function estimation procedure, we applied it to determine the dark matter particle-halo cross-correlation function, and found close agreement with an alternate particle-based estimation code from \cite{sinha_corrfunc_2020} across the scales relevant for our current study, $r_\perp \sim 0.4-50  $ Mpc/$h$.

We will also consider $\langle |\mathrm{RM}| \times \mathrm{g} \rangle$ cross-correlations for comparison. This statistic -- or slight variations of it -- has been considered in previous studies, both in simulations \citep{stasyszyn_measuring_2010} and observations \cite{amaral_constraints_2021}. The latter study placed an upper bound on a closely related quantity from observational data. 

Note that the cross-correlation estimators used in \cite{stasyszyn_measuring_2010} and \cite{amaral_constraints_2021} compute the average galaxy abundance in radial bins around RM sources, weighted by the deviation of each $\rm |RM|$ value from the global mean. This approach is closely related to our method, which measures the excess $|{\rm RM}|$ (or ${\rm RM}^2$) in bins around galaxies. The two estimators are mathematically equivalent when applied to a continuous, smoothed RM field, under the assumption that the RM and galaxy density fields are statistically homogeneous and isotropic. As discussed earlier, however, a discrete RM estimator probes the RM field on smaller scales than in our map-based approach.

\section{Simulated Cross-Correlation Signals and Their Interpretation} \label{sec:results}

\subsection{The Simulated $\langle \mathrm{RM}^2 \times \mathrm{g} \rangle$ and $\langle |\mathrm{RM}| \times \mathrm{g} \rangle$ Signals}

\begin{figure*}[ht!]
    \centering
    \includegraphics[width=0.49\linewidth]{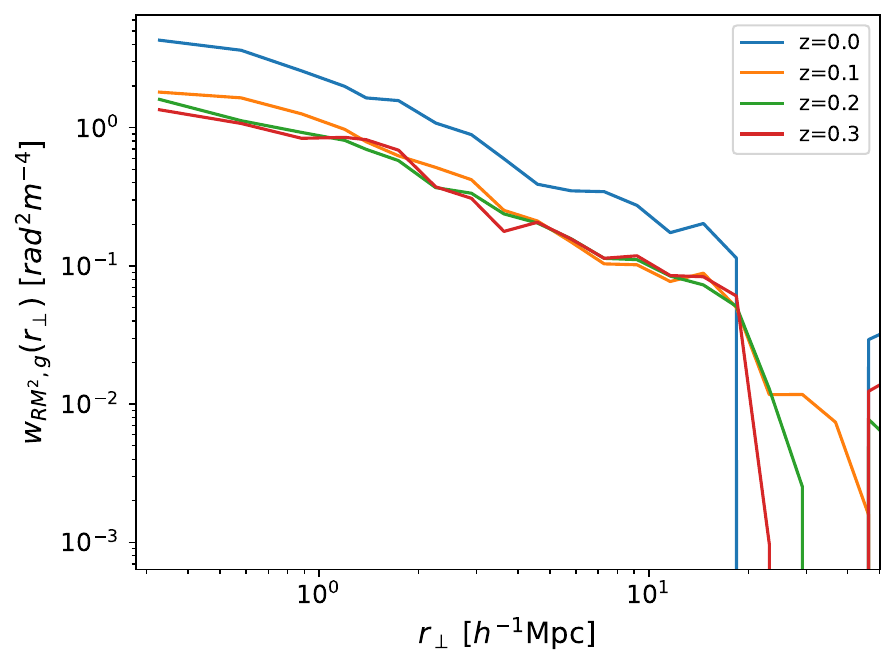}
    \includegraphics[width=0.49\linewidth]{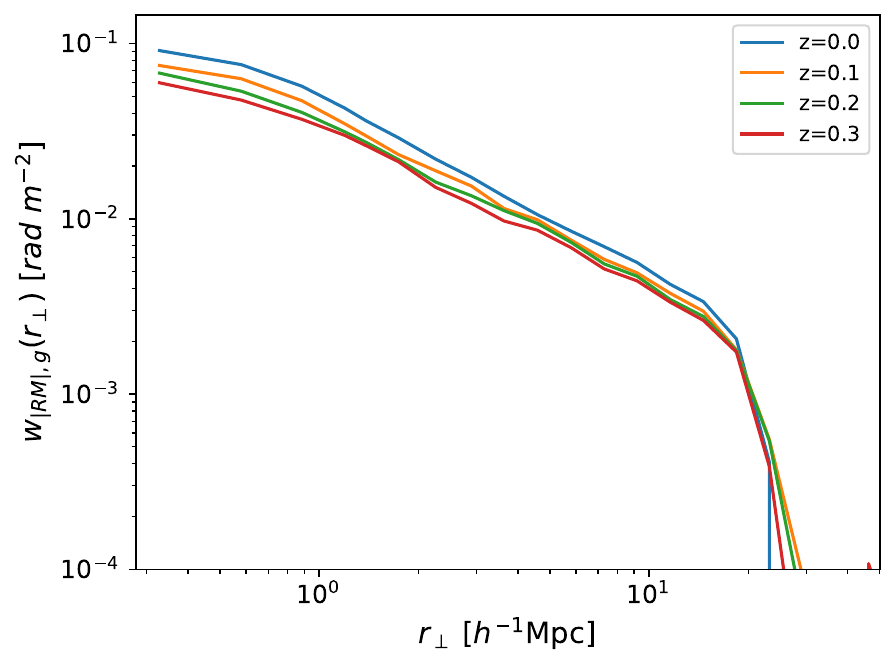}
    \includegraphics[width=0.49\linewidth]{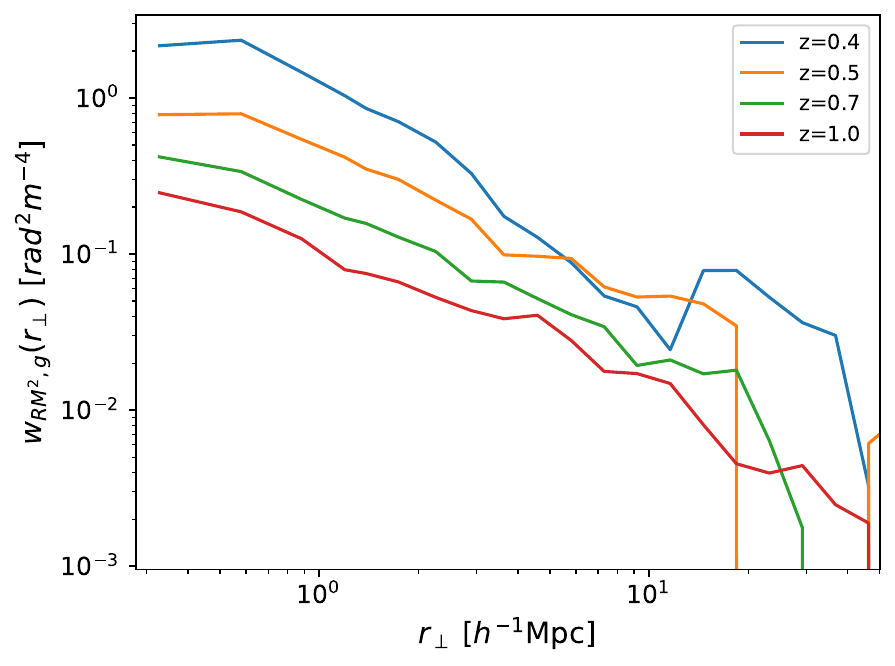}
    \includegraphics[width=0.49\linewidth]{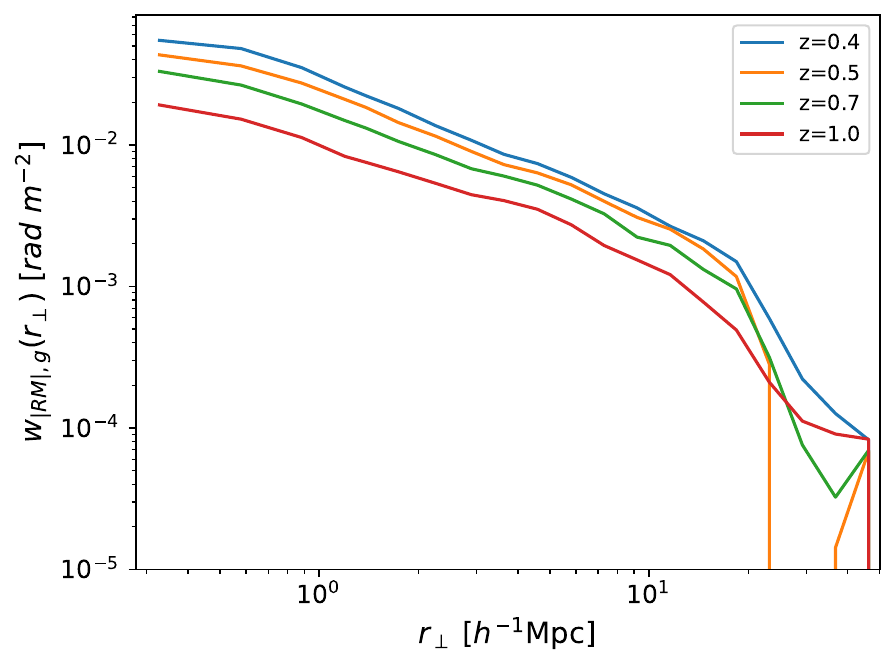}
    \includegraphics[width=0.49\linewidth]{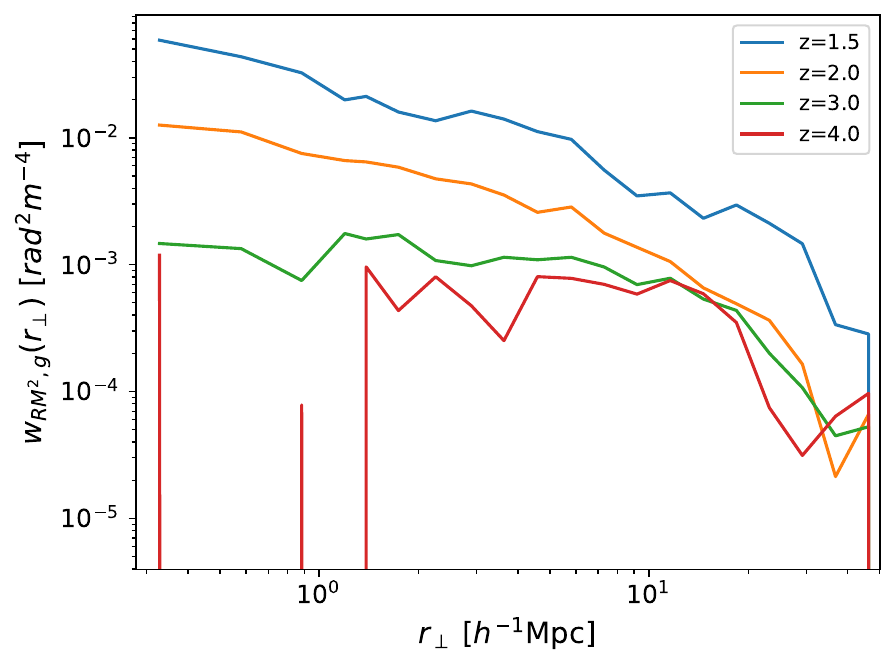}
    \includegraphics[width=0.49\linewidth]{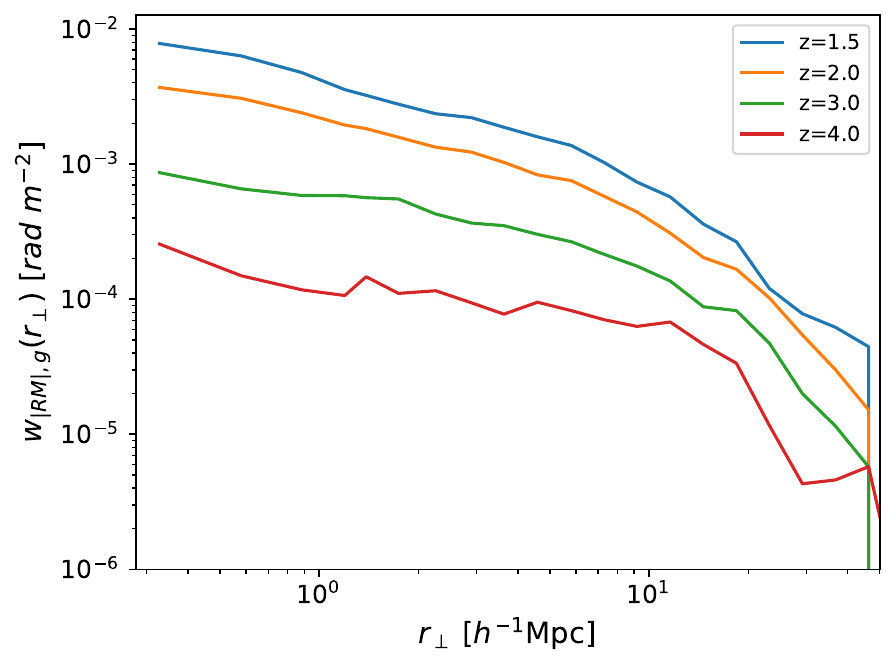}
    \caption{Two-point cross-correlation functions between the RM and halo overdensity fields, computed using two estimators: $w_{\mathrm{RM}^2,{\rm g}}(r_\perp;z)$ ({\em left panels}) and $w_{|\mathrm{RM}|,{\rm g}}(r_\perp;z)$ ({\em right panels}). Here we show coeval cross-correlations in narrow redshift bins (see text) from $z=0$ to $z=4$. Note that the scales considered here lie beyond the virial radius of most of the halos included in the stacks, and so the signal here reflects the large-scale correlated excess ${\rm RM}^2$ (or $|\mathrm{RM}|$ around the stacked halos. The cross-correlations increase strongly with decreasing redshift, as we will discuss. The two estimators show broadly similar scale dependence and redshift evolution trends.     } 
    \label{fig:rm-gal}
\end{figure*}

In general, the observed RM towards a background source consists of three distinct components:
\begin{equation}
    \mathrm{RM}_{\mathrm{obs}} = \mathrm{RM}_{\mathrm{in}} + \mathrm{RM}_{\mathrm{path}} + \mathrm{GRM},
    \label{eq:obs}
\end{equation}
where $\mathrm{RM}_{\mathrm{in}}$ is the intrinsic contribution from magnetized plasma within the source, $\mathrm{GRM}$ is the galactic foreground contribution from the Milky Way, and $\mathrm{RM}_{\mathrm{path}}$ is the contribution from other structures along the path of propagation. As alluded to earlier, we can statistically extract the portion of $\mathrm{RM}_{\mathrm{path}}$ at redshifts close to those of 
tracer galaxies, as a function of the tracer redshift. The remaining components, $\mathrm{RM}_{\mathrm{in}}$ and $\mathrm{GRM}$, arise from regions which are physically distant from the tracer galaxies, and should
therefore be uncorrelated with the galaxy survey and act only as noise. The cross-correlation statistics hence
isolate the RM contributions at different redshifts. 

Figure~\ref{fig:rm-gal} shows cross-correlations between each of the RM$^2$ and $|\mathrm{RM}|$ fields with the halo distribution in \textsc{Illustris-TNG} for narrow halo redshift bins from $z=0-4$. As anticipated, the simulations show fairly strong excess RM$^2$ and $|\mathrm{RM}|$ signals close to the halo centers. In the regimes considered here, these signals mainly reflect the contribution from the magnetized plasma in correlated neighboring systems, rather than from material within the central halo itself. This is the case because we consider $r_\perp \gtrsim 0.4$ Mpc/$h$ here, which is beyond the virial radius of most of the halos in the stack. Nevertheless, each estimator picks up a clear excess RM$^2$ or $|\mathrm{RM}|$ signal from correlated neighboring halos. That is, our statistic is measuring the correlated excess RM$^2$ or $|\mathrm{RM}|$ signal from the surrounding cosmic web. 

In principle, this large-scale signal contains contributions from the magnetic fields in neighboring halos, filaments, and other environments in the surrounding large-scale structure. In practice, in \textsc{Illustris-TNG}, the dominant contributions to the average correlated RM$^2$ signal come from regions within neighboring halos, as we will discuss. We emphasize that our goal here is not to isolate volume-filling intergalactic magnetic fields, which would require additional analysis steps to mitigate the substantial RM$^2$ contributions from neighboring halos in the stacked signal. Instead, our objective is to constrain the redshift evolution of the overall magnetic energy density, even if this signal is dominated by magnetized plasma within dark matter halos.

The cross-correlation signals decline with increasing $r_\perp$: this is natural given that correlations between the central and neighboring halo positions decrease with growing spatial separation. We will sharpen this statement for the case of $w_{\mathrm{RM}^2, \mathrm{g}}(r_\perp;z)$ in the following subsection. 
At large scales, both estimators become noisy and suffer from finite boxsize effects around $r_\perp \gtrsim 20$ Mpc/$h$ or so (see e.g. \cite{springel_first_2018} where similar boxsize effects can be observed in estimates of the projected galaxy correlation functions from \textsc{Illustris-TNG}). 

Another notable feature is the strong redshift evolution in both $w_{\mathrm{RM^2,g}}(r_\perp;z)$ and $w_{|\mathrm{RM}|,\mathrm{g}}(r_\perp;z)$. In each case, the signals increase strongly towards low redshifts, with the signal strength growing by around 3-4 orders of magnitude from $z \sim 4$ to $z \sim 0$. As we will discuss and quantify, this arises largely because of the progressive amplification of magnetic field strengths across cosmic time. A key point here is that measurements of $w_{\mathrm{RM^2,g}}$, and possibly $w_{|\mathrm{RM}|, \mathrm{g}}$, as a function of redshift, offer a relatively clean means to extract the evolution of large-scale magnetic fields empirically. 

The $w_{\mathrm{RM}^2, \mathrm{g}}(r_\perp;z)$ signal also shows a non-uniform evolution across redshift. Specifically, it evolves little between $z=0.1-0.3$, while the amplitude subsequently increases significantly towards $z=0$. We attribute the lack of evolution across $z=0.3-0.1$ to the saturation of turbulent dynamo amplification processes, as expected once the magnetic energy density reaches a fixed fraction of the turbulent kinetic energy density \cite{garcia_magnetization_2021}.
The subsequent increase from $z=0.1$ to $z=0$ arises because, as discussed in the following subsection, the signal amplitude is determined by the electron-density–weighted magnetic field, and the electron distribution in Illustris-TNG exhibits enhanced small-scale clumpiness at late times.

Although both estimators show broadly similar redshift and scale dependence, there are some important differences between the two correlation functions. First, $w_{\mathrm{RM^2,g}}$ is easier to connect to the statistical properties of the underlying fields involved. That is, it can be more readily connected to the power spectra and bispectra of the magnetic field, electron density, and galaxy/halo distribution fluctuations. Second, in Section~\ref{sec:obs_noise} we will show that $w_{|\mathrm{RM}|, \mathrm{g}}$ suffers from noise bias, while $w_{\mathrm{RM^2,g}}$ is immune to noise bias. On the other hand, a potential advantage of $w_{|\mathrm{RM}|, \mathrm{g}}$ is that it is less dominated by rare high RM pixels, which are amplified by the quadratic weighting in $w_{\mathrm{RM^2,g}}$.
This makes $w_{\mathrm{RM^2,g}}$ especially sensitive to the tails in the joint probability distribution of magnetic field strength and electron density. For example, the {\em  mean} electron-density--weighted magnetic field is much larger than the {\em median} one. 
Since $w_{|\mathrm{RM}|, \mathrm{g}}$ involves only a linear weighting of the RM data, it should be less dominated by the rare high RM pixels. In practice, we advocate measuring both statistics, as their joint measurement and interpretation should help ensure the robustness of our conclusions. 

\subsection{Analytic Model for \texorpdfstring{$\langle \mathrm{RM}^2 \times \mathrm{g} \rangle$}{<RM² × g>} Cross-Correlations}

To help interpret the simulation results for $\langle \mathrm{RM}^2 \times \mathrm{g} \rangle$, we develop
an analytic approximation. Specifically, in Appendix \ref{app:a} we argue that the triangle power spectrum of Eq.~\ref{eq:triangle_power} may be
approximated as:
\begin{equation}
 \label{eq:tofk_refine}
 \mathcal{T}(k) \approx 2 K P_{e,g}(k) \int \frac{d^2q_\perp}{(2 \pi)^2} P_{\tilde{B}}(q_\parallel=0,q_\perp),
 \end{equation}
where \(K\) is a redshift-independent calibration factor determined from the simulations, \(P_{e,g}(k)\) is the electron–galaxy cross-power, and \(P_{\tilde B}(q)\) is the 3D power spectrum of the electron-density--weighted line-of-sight magnetic field \(\tilde B_\parallel \equiv [1+\delta_e] B_\parallel\). The electron-density--weighted magnetic field power is evaluated for transverse Fourier modes, with $q_\parallel=0$, owing to the projection involved. The triangle power spectrum $\mathcal{T}(k)$ hence includes an integral of the $\tilde{B}_\parallel$ power spectrum over $q_\perp$. 

As detailed in Appendix \ref{app:a}, this approximation comes from a lowest-order expansion of the bispectrum in Eq.~\ref{eq:triangle_power}, neglecting higher-order terms and cross-correlations between electron and magnetic field fluctuations (which have distinct coherence scales).  However, it includes a heuristic step which we do not attempt to rigorously justify. In particular, our approximation involves replacing $P_B$ with $P_{\tilde{B}}$ (i.e., swapping the magnetic field power spectrum, which arises in the lowest-order expansion, with its electron-density--weighted counterpart) and introducing $K$. These steps are intended to approximately capture electron-weighting and non-linear (beyond leading order) effects, but lack rigorous justification. Although it would be interesting to investigate this further in future work, the approximate formula provides physical intuition and matches the simulation results fairly well.  

Substituting Eq.~\ref{eq:tofk_refine} into Eq.~\ref{eq:wag_triangle} yields
\begin{equation}
w_{\mathrm{RM^2,g}}(r_\perp = \chi \theta; z) = 2K W_{\mathrm{RM}}^2  \int \frac{dk}{2\pi} k J_0\left[k (\chi \theta)\right] P_{e,g}(k; z) \times \int \frac{d^2 q_\perp}{(2\pi)^2} P_{\tilde{B}}(q_\parallel=0, q_\perp; z),
\label{eq:wrm_approx}
\end{equation}
where \(W_{\mathrm{RM}}\) is the RM prefactor defined earlier in Eq.~\ref{eq:rm_chi}.
This expression highlights that the \emph{shape} of the RM$^2$–halo correlation is governed primarily by how electrons cluster around galaxies or halos, \(P_{e,g}(k;z)\), while the \emph{normalization} is set by the projected strength of the density–weighted magnetic field through the term \(\propto \int d^2q_\perp P_{\tilde B}\).

\begin{figure}[ht!]
    \centering
    \includegraphics[width=0.6\linewidth]{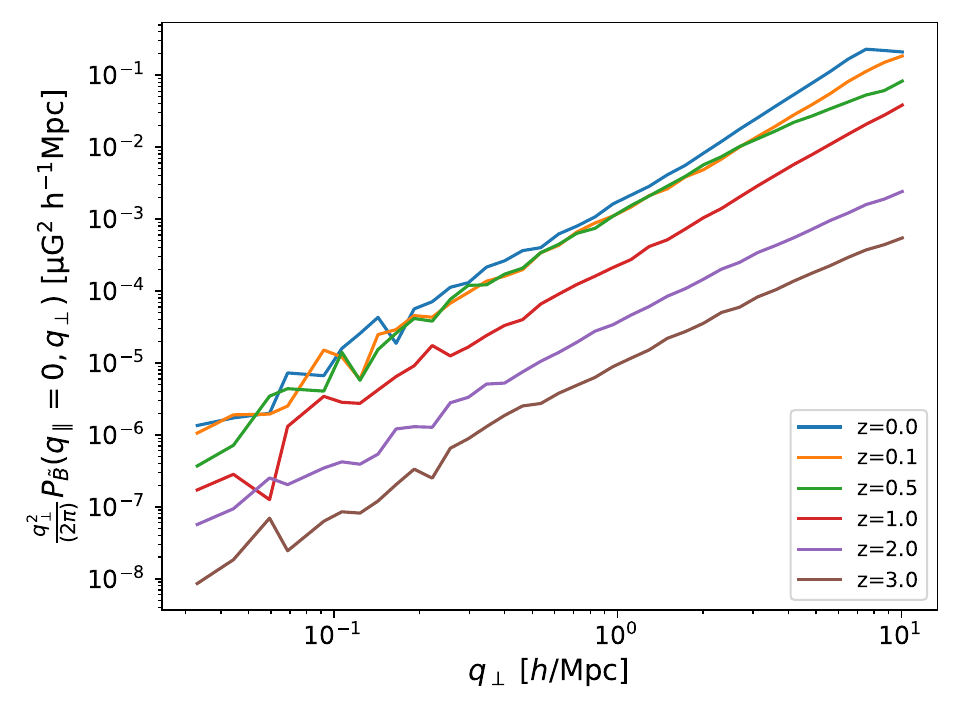}
    \caption{Power spectrum of the electron-density--weighted line-of-sight magnetic field as a function of redshift.
    Here we take only the $q_\parallel=0$ modes of this field, because line-of-sight Fourier modes are suppressed in projection. The quantity plotted shows the contribution to the electron-weighted magnetic field strength per $d\,\rm{ln} \,q_\perp$. The integral of this power spectrum, over all $q_\perp$, largely governs the amplitude of $w_{\mathrm{RM}^2,{\rm g}}(r_\perp;z)$. In \textsc{Illustris-TNG} the electron-density--weighted magnetic field strength increases strongly towards low redshift. 
    }
    \label{fig:Pb}
\end{figure}

In order to get a feel for the important ingredients here, we first plot the power spectrum of the electron-density--weighted line-of-sight magnetic field in Fig.~\ref{fig:Pb}. Here we show the contribution per logarithmic interval in $q_\perp$, i.e., $\propto q^2_\perp P_{\tilde{B}}(q_\parallel=0, q_\perp)$. This helps to understand which Fourier modes dominate the integral in Eq.~\ref{eq:tofk_refine} and Eq.~\ref{eq:wrm_approx}.
Note that the units of \(\int d^2q_\perp \, P_{\tilde B}(q_\parallel=0,q_\perp)\) are $\mu G^2 \, {\rm Mpc}/h$. First,
the figure shows that the power spectrum is close to a white-noise spectrum: the term 
$\propto q^2_\perp P_{\tilde{B}}(q_\parallel=0, q_\perp)$ increases just a little more steeply towards small-scales than the white-noise expectation, $q^2_\perp P_{\tilde{B}}(q_\parallel=0, q_\perp) \propto q^2_\perp$. This shows that the electron-density--weighted magnetic field strength fluctuations are dominated by small-scale Fourier modes. Second, this power spectrum grows rapidly towards decreasing redshifts, with the overall amplitude increasing by about three orders of magnitude from $z \sim 3$ to $z \sim 0$.

\begin{figure}
    \centering
    \includegraphics[width=0.49\linewidth]{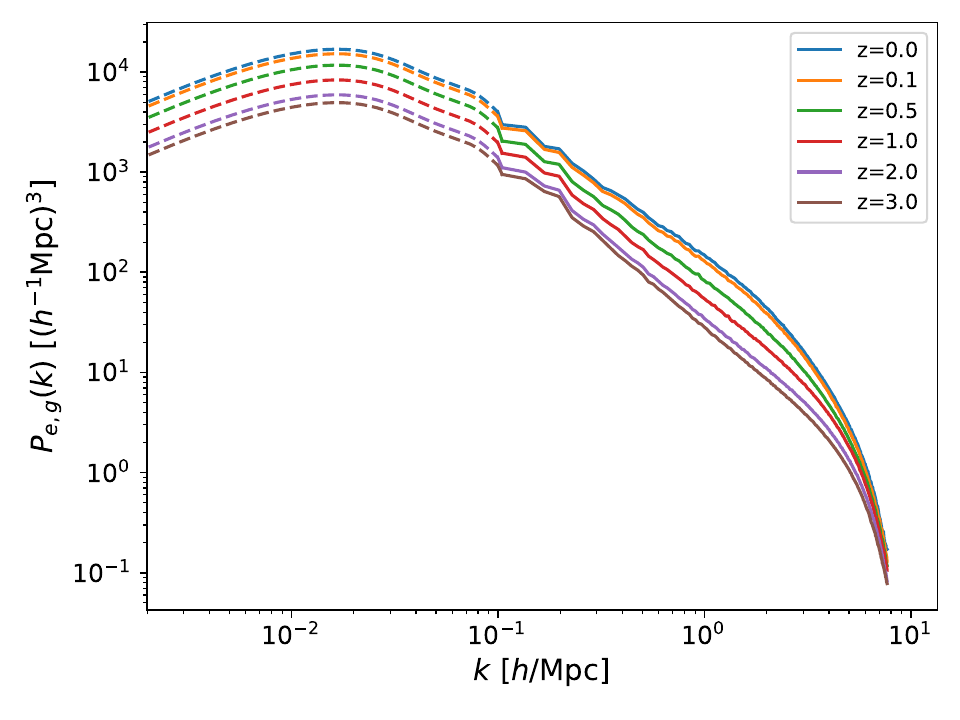}
    \includegraphics[width=0.49\linewidth]{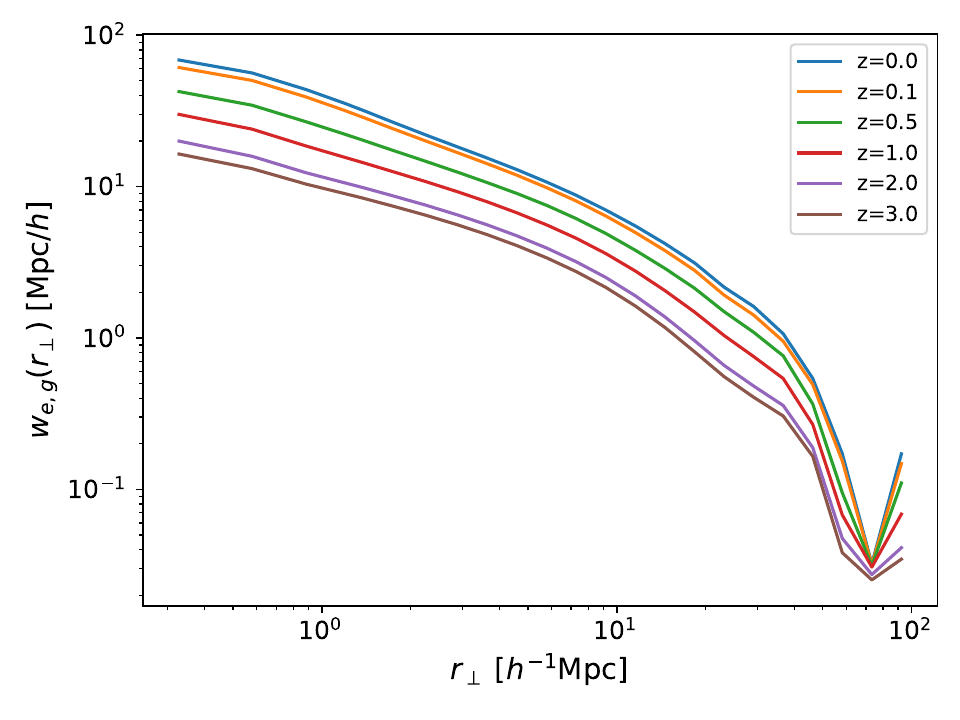}
    \caption{
    The (3D) electron density–halo cross-power spectrum ({\em left panel}) and the real-space projected electron-halo two-point cross-correlation function ({\em right panel}).  These functions characterize spatial correlations between the free electron and halo distributions, and largely determine the large-scale shape of the $w_{\mathrm{RM}^2,{\rm g}}(r_\perp;z)$ correlation function. 
    We extrapolate the simulated 3D electron density-halo cross-power spectrum to large scales (beyond the fundamental mode of the simulation box) assuming a linear biasing model (dashed lines). The projected electron-halo cross-correlation function is computed based on the cross-power in the left panel, and the linear theory extrapolation to low $k$.  The amplitude of the projected electron-halo correlation function increases towards low redshift.  
    }
    \label{fig:Peg}
\end{figure}

Next, we consider $P_{e,g}(k)$ in Fig.~\ref{fig:Peg}. The left-hand panel shows the 3D electron-halo cross-power spectrum as a function of wavenumber and redshift. The dashed lines show extrapolations to low $k$ based on a linear biasing model:  on large-scales the cross-power is assumed to follow $P_{e,g}(k) = b_e b_g P_{\rm lin}(k)$, where the linear theory power spectrum is computed using the Eisenstein \& Hu transfer function \citep{Eisenstein_1998}, as implemented in the \texttt{Colossus} package \citep{diemer_colossus_2018}. The factor $b_e b_g$ is adjusted to match the simulation results at low $k$. Using the entire \textsc{Illustris-TNG} halo catalog we find, for example, $b_e b_g = 2$ at $z=3$ and $b_e b_g = 0.72$ at $z=0$.
This linear biasing model allows us to extrapolate predictions to length-scales beyond those that are well-captured in the TNG300 series \textsc{Illustris-TNG} simulation suite. The right-hand panel shows the resulting projected electron-halo cross-correlation function, computed as the first integral in Eq.~\ref{eq:wrm_approx}. The linear biasing model allows reliable predictions at large $r_\perp$. Although $w_{e,g}(r_\perp; z)$ shows some growth towards low $z$, the evolution is much milder than in the projected electron-density--weighted magnetic field strength. Hence, our approximate expression (Eq.~\ref{eq:wrm_approx}) suggests that the redshift evolution in the amplitude of $w_{\mathrm{RM^2,g}}(r_\perp;z)$ is primarily dictated by growth in the electron-density--weighted magnetic field strength. In the approximate formula, the shape of the RM$^2$-galaxy correlation function is, however, governed by $w_{e,g}(r_\perp;z)$. Here, the relevant scale dependence in \textsc{Illustris-TNG} can be discerned by inspecting the right-hand panel of Fig.~\ref{fig:Peg}.
The results at each redshift show a similar scale dependence. 
These curves can also potentially be determined empirically via other measurements, including fast radio burst (FRB) dispersion measure-galaxy cross-correlations (DM-galaxy) \citep{Leung25,Madhavacheril19,McQuinn14} and kSZ-galaxy cross-correlations \citep{Hadzhiyska25}. 

Note that although there are good prospects for empirical constraints on the projected electron-galaxy cross-correlation function, $w_{e,g}(r_\perp;z)$,
using our methodology to constrain magnetic fields does not hinge crucially on obtaining such measurements. On the scales of interest, $w_{e,g}(r_\perp;z)$ is well-described by a linear biasing model, where the scale dependence follows the well-understood matter correlation function. In this regime, we only need to model the effective bias factor of the electron distribution. This can be modeled fairly reliably, especially relative to the much larger uncertainties in the magnetic field strengths we aim to constrain.

\begin{figure}
    \centering
    \includegraphics[width=0.6\linewidth]{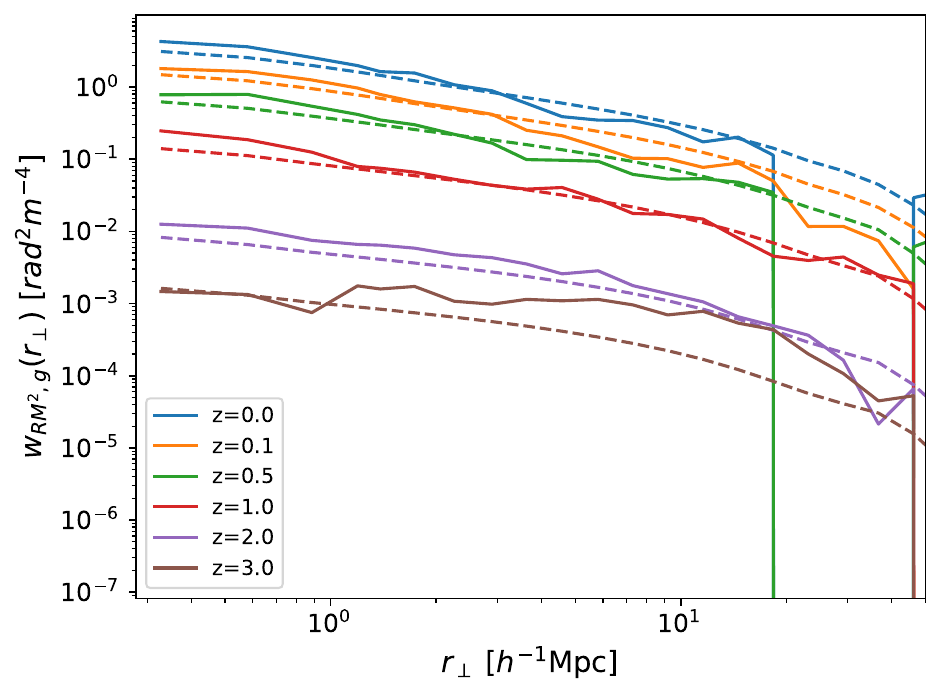}
    \caption{Comparison between the simulated RM$^2$-halo cross-correlation function and the analytical approximation of Equation~\ref{eq:wrm_approx}, evaluated at various redshifts. The solid lines show measurements from \textsc{Illustris-TNG}, while the dashed lines are the analytic results. The agreement is generally good, demonstrating that the analytic model captures much of the redshift evolution and scale dependence of the simulated signals. Note that, in agreement with the model, the amplitude of the correlation function in the simulations drops by roughly three orders of magnitude from $z=0$ to $z=3$. 
    The analytic model is most successful at $r_\perp \gtrsim 1$ Mpc/$h$.  
     At large separations, $r_\perp \gtrsim 10$ Mpc/$h$, we expect the model to be reasonably accurate but the simulation measurements are too noisy to provide a sharp test. The approximate model appears less successful at the highest redshift shown, $z =3$. 
    }
    \label{fig:approx_model}
\end{figure}

Figure~\ref{fig:approx_model} compares the \textsc{Illustris-TNG} simulation results for $w_{\mathrm{RM^2,g}}$ against the analytic approximation of Eq.~\ref{eq:wrm_approx}. The constant factor $K$ in the equation has been adjusted to match the simulation results on intermediate scales around $1 \, \mathrm{Mpc}/h \lesssim r_\perp \lesssim 10 \, \mathrm{Mpc}/h$, yielding $K=3$. The analytic approximation provides a relatively good description of the simulation results, keeping in mind that the simulation measurements become noisy and suffer from finite-size box effects at $r_\perp \gtrsim 10-20$ Mpc/$h$ or so. The amplitude, redshift evolution, and scale dependence are broadly reproduced by the simple approximate formula, even though the strength of the correlation grows by approximately three orders of magnitude from $z \sim 3$ to $z \sim 0$. The model is generally less successful on the smallest scales ($r_\perp \lesssim 1$ Mpc/$h$), where it is unlikely to fully capture non-linear effects.  

Otherwise, the most notable discrepancy between the simulation results and the model is at $z=3$, where the cross-correlation exhibits a flatter scale dependence than expected from the analytic formula. The origin of this behavior remains unclear and requires further investigation. However, note that the observational prospects at such high redshifts are currently limited 
since this requires RM measurements towards sources at $z \gtrsim 3$ and relatively high number density galaxy catalogs for cross-correlations.

The comparison between the simulation and analytic results supports the notion that the redshift evolution in the RM$^2$-halo cross-correlation is primarily driven by the growth of the projected electron-density--weighted magnetic field strength across cosmic time. That is, the amplitude of the correlation function mainly traces $\int \frac{d^2 q_\perp}{(2 \pi)^2} P_{\tilde{B}}(q_\parallel=0,q_\perp;z)$ (Eq.~\ref{eq:wrm_approx}). Thus, measurements of $w_{\mathrm{RM^2,g}}(r_\perp;z)$ across redshift effectively trace the \textit{cosmic evolution of magnetic energy density}, provided the electron-galaxy cross-correlation, $w_{e,\mathrm{g}}(r_\perp;z)$, can be modeled or constrained independently from FRB and/or kSZ cross-correlation measurements. Relating the above projected field integral to the volume-averaged magnetic energy density requires modeling the coherence length of $\tilde{B}$, since the coherence scale links projected rms fluctuations to their full 3D values. One must also account for the electron-density weighting inherent in $\tilde{B}$. Despite these caveats, this approach offers a relatively direct route towards a census of cosmic magnetic energy density across cosmic time. The consistency of the simulated correlation function with our analytic model is encouraging. It suggests that we will be able to robustly interpret future measurements of $\langle \mathrm{RM}^2 \times \mathrm{g} \rangle$, without relying on the detailed assumptions inherent in the \textsc{Illustris-TNG} models.

\subsection{On the Redshift Evolution of Magnetic Field Strength in \textsc{Illustris-TNG}}

\begin{figure*}
    \centering
    \includegraphics[width=0.41\linewidth]{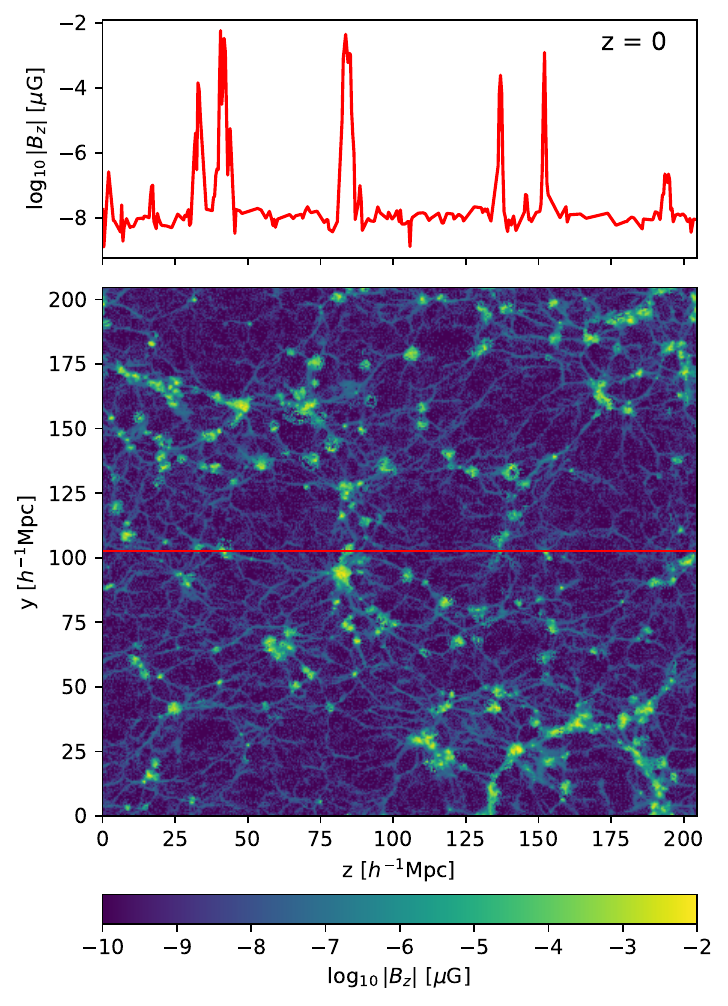}
    \includegraphics[width=0.41\linewidth]{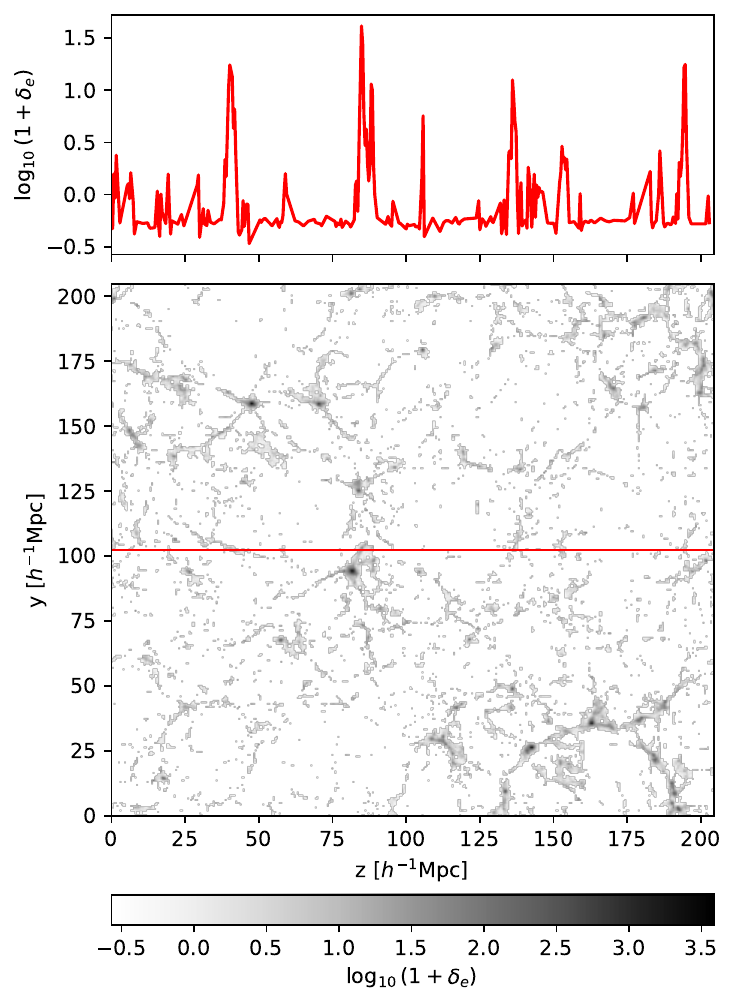}
    \includegraphics[width=0.41\linewidth]{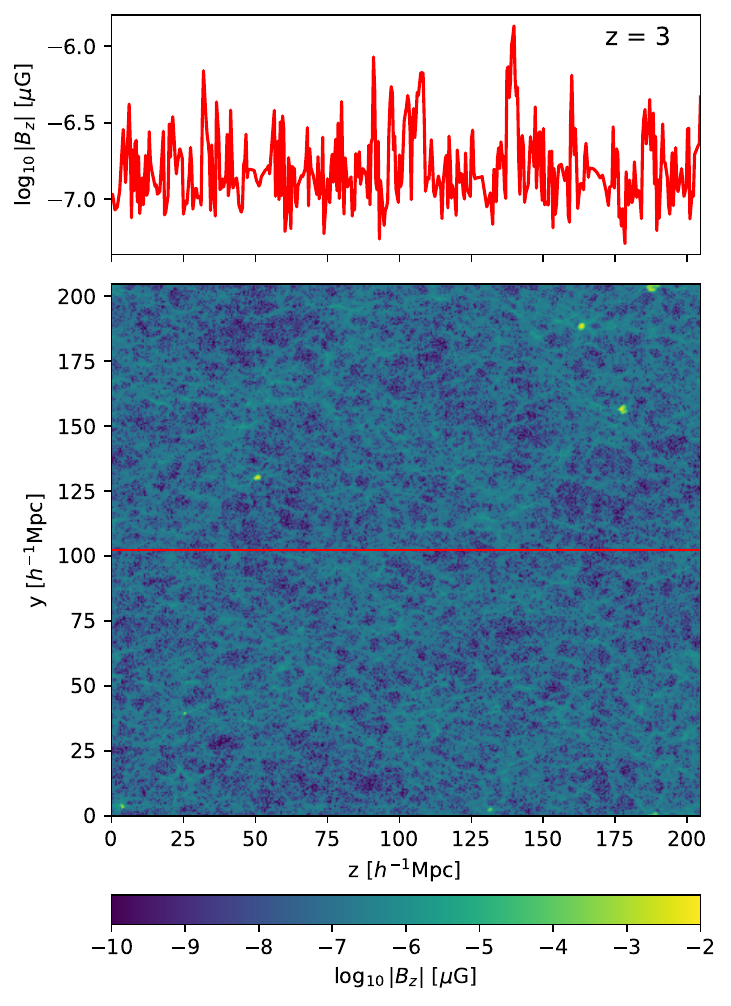}
    \includegraphics[width=0.41\linewidth]{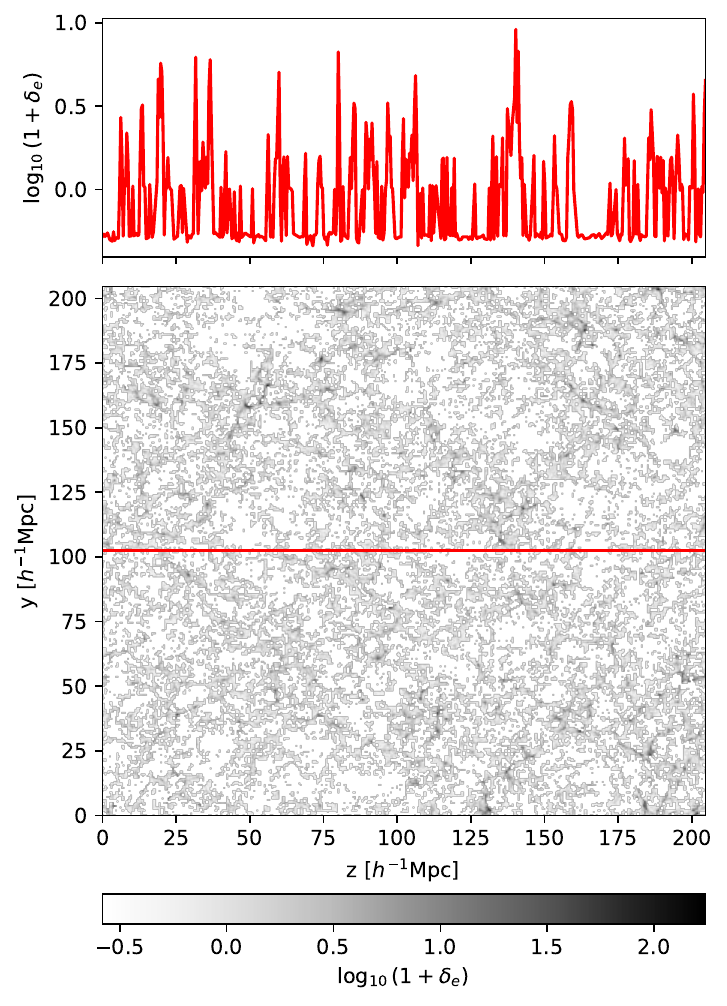}
    \caption{
    Line-of-sight magnetic field strength ({\em left column}) and electron density ({\em right column}) in TNG300 over a region of size $(205~\mathrm{Mpc}/h)^2 \times 0.41~\mathrm{Mpc}/h$, at redshifts $z = 0$ ({\em top row}) and $z = 3$ ({\em bottom row}). The {\em upper panels} show the magnetic field and electron overdensity along an example line of sight (indicated in red in the {\em lower panels}). At $z=0$ the magnetic field strength is strongly enhanced in overdense regions, partly
    owing to magnetized outflows in the simulations, which are visible as ``bubble''-like structures in the {\em top left} panel. At $z=3$, both the electron and magnetic field distributions are more uniform. The magnetized bubbles at this redshift are smaller and fill less of the simulation volume, while magnetic dynamo and other amplification processes have been underway for less time. In \textsc{Illustris-TNG}, the overall amplitude of electron-density--weighted magnetic field strength increases strongly with decreasing redshift owing to these magnetized outflows and amplification processes. The amplitude of the $\langle {\rm RM}^2 \times {\rm g} \rangle$ correlation versus redshift largely traces this evolution. Inspired by a related figure in \cite{aramburo-garcia_magnetization_2021}. 
}
    \label{fig:los}
\end{figure*}

Next we turn to consider the strong increase in the electron-density--weighted magnetic field towards low redshifts observed in our calculations from the \textsc{Illustris-TNG} simulations. 
Note that in low-density regions, magnetic-flux--freezing in the ionized IGM causes the magnetic fields in these regions to scale as $(1+z)^2$, leading to a decline over time. In contrast, dynamo amplification processes lead to significant enhancements within collapsed halos, which grow over time \citep{marinacci_large-scale_2015, aramburo-garcia_magnetization_2021, aramburo-garcia_revision_2022}. The amplified magnetic fields also spread into the surrounding gas, driven by AGN and galactic outflows, which produce bubbles of magnetized plasma that can extend beyond the virial radius of the host dark matter halos \citep{aramburo-garcia_magnetization_2021}. 

In order to obtain a further qualitative feel for which structures shape the $\langle \mathrm{RM}^2 \times \mathrm{g} \rangle$ signal, Figure~\ref{fig:los} shows visualizations across example slices through \textsc{Illustris-TNG} simulation snapshots, inspired by a similar figure in \cite{aramburo-garcia_magnetization_2021}. Here, we extract the LoS magnetic field strengths and electron overdensities across representative sightlines at $z=0$ and $z=3$. At low redshifts, the electron density contrasts are sharpened and amplified relative to the higher redshift case, a natural consequence of gravitational instability. More striking is the enormous contrast in the $z \sim 0$ magnetic field strengths, which vary by six orders of magnitude along the example LoS shown in the figure. 
At $z=3$, the corresponding spread in field strengths is only about a factor of ten. 
Note that the smallest values of the magnetic field actually drop from $z=3$ to $z=0$: this is a consequence of magnetic flux-freezing in the low density IGM.
The $z=0$ magnetic field strength in the middle-left panel of the figure shows a partly web-like and partly bubble-like topology, while the $z = 3$ field shows a more diffuse and purely web-like structure. Only a handful of tiny magnetized bubbles are visible in the slice at $z \sim 3$. These results illustrate the progressive amplification of magnetic fields in overdense regions, along with the growth of magnetized bubbles over time. 

For present purposes, the key point is that our $\langle \mathrm{RM}^2 \times \mathrm{g} \rangle$ statistic is dominated by highly-magnetized overdense environments. The electron-density--weighted magnetic field strength then increases sharply towards low-$z$, driven by the amplification processes occurring in overdense regions.
Furthermore, note that our correlation-function estimator determines the {\em mean} value of the RM$^2$ field in an annulus around a halo or galaxy center. The highly non-Gaussian magnetic field strength and electron overdensity fields, evident in Figures~\ref{fig:data} and \ref{fig:los}, imply that the mean values of RM$^2$ may greatly exceed the {\em median} RM$^2$ values. In any case, the power of our method is that it delivers a census of the projected electron-density-weighted field strength at a given redshift. This should follow without relying on a particular cosmological MHD model, at least to the extent that our simulation-calibrated analytic formula (Eq.~\ref{eq:wrm_approx}) holds generally. In order to further interpret the inferred projected electron-density--weighted field strengths, however, it will be important to compare with \textsc{Illustris-TNG} and other MHD simulations. 
Additional insights will likely be gained by measuring other statistical properties of the RM fields, including $\langle |\mathrm{RM}| \times \mathrm{g} \rangle$, which should be less dominated by highly overdense regions. 

\subsection{The Effect of the RM Smoothing-Scale}
\label{sec:fiducial_rm}

\begin{figure}
    \centering
    \includegraphics[width=0.6\linewidth]{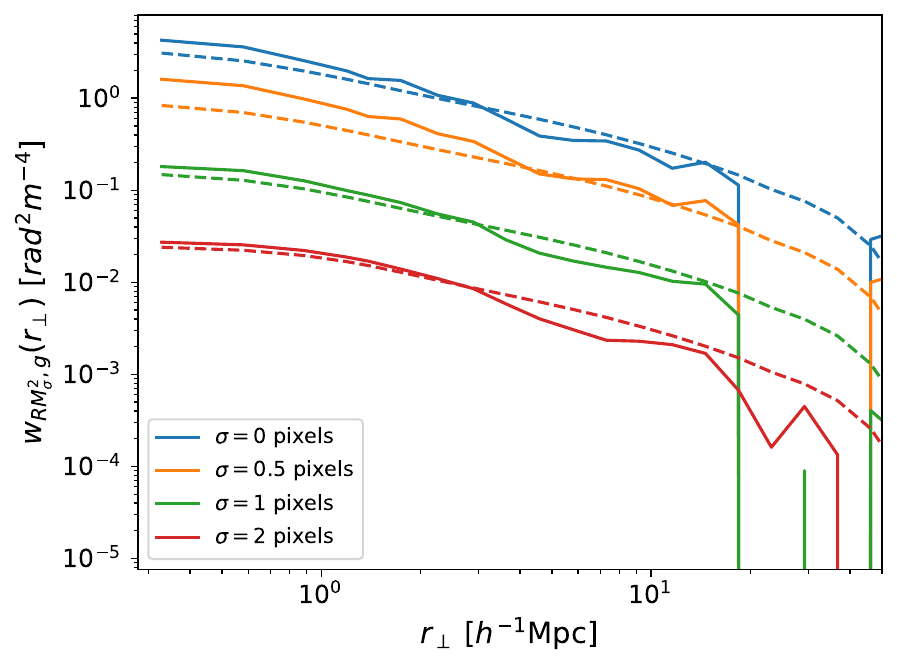}
    \caption{Simulated RM$^2$–halo cross-correlation function and the analytical approximation from Equation~\ref{eq:wrm_f}, evaluated at redshift $z = 0$ with various Gaussian smoothing scales, spanning from $\sigma = 0.5$ pixels $=0.21$ Mpc/$h$ out to $\sigma = 2$ pixels $= 0.82$ Mpc/$h$.
    Solid lines show measurements from \textsc{Illustris-TNG}, with smoothing applied to simulate coarser RM grids. The cross-correlation is a strong function of smoothing scale; this is expected based on the $\tilde{B}$ power spectrum of Figure~\ref{fig:Pb} which shows that high $q_\perp$ fluctuations are strong. In practice, this implies that the signal is sensitive to the precise smoothing scale of the measurements.   
}
    \label{fig:filter_wrm}
\end{figure}

Since our $\langle \mathrm{RM}^2 \times \mathrm{g} \rangle$ statistic involves the {\em squared} field, it is sensitive to the precise smoothing scale of the RM measurements. This is especially the case since the $\tilde{B}$ power spectrum includes strong high-$q_\perp$ fluctuations (see Figure~\ref{fig:Pb}).  In practice, the RM measurements will be beam-smoothed at the angular resolution of the observations, and additional filtering may be applied when constructing RM maps or during foreground cleaning steps. 

Our mock RM fields are implicitly smoothed at the resolution of the mesh we use to grid the magnetic field and electron density fields before taking their product and projecting along the line of sight. Note that this is a 3D smoothing and this is not precisely the same as the angular smoothing that will be applied to the RM measurements from current and future surveys. Nevertheless, we note that the cell size of our 3D mesh is 0.41 Mpc/$h$. This is at least broadly comparable to the expected mean separation between the RM sources in the case of upcoming SKA data. Ultimately, the SKA is expected to measure roughly $\sim 2 \times 10^7$ RMs across nearly the full sky \citep{gaensler_origin_2004}.  This corresponds to a mean source separation of $160''$ or a comoving spacing of 0.46 Mpc/$h$ at a typical foreground galaxy redshift of $z=0.2$. This is similar to the size of our 3D mesh. 
Hence, if we consider measurements from an SKA RM map, with a resolution governed by the RM source separation, our gridding procedure should broadly mimic the expected smoothing in the data. In order to make a detailed comparison with upcoming data, however, it will be important to employ a finer 3D mesh and to closely match the angular resolution of the RM observations in our mock data. 

In principle, one should also smooth over the transverse length scale of each polarized RM source. In practice, however, the typical angular sizes of the polarized sources used for RM measurements are $\lesssim 1-20''$ (e.g. \cite{Hales14}), which are much smaller than the effective smoothing scale considered here. 

We emphasize the distinction between discrete and map-based estimators, mentioned earlier. For point-like or highly compact background sources, individual RM measurements correspond to pencil-beam line integrals and are sensitive to small-scale IGM structure, potentially below the resolution of cosmological MHD simulations. In contrast, our map-based approach applies an explicit smoothing to the RM field prior to squaring, introducing a well-defined transverse averaging scale that suppresses sensitivity to sub-resolution structure.

Consequently, in this regime, where the intrinsic source sizes remain small compared to the pixelization or beam window, the additional convolution with the intrinsic source profile introduces only a minor modification to the smoothing filters of Eq.~\ref{eq:A_sigma_conv} (see below) and does not significantly impact our predicted signals. In the case of higher angular resolution data, however, the finite and variable angular sizes of polarized sources will become increasingly important. In that case, a practical mitigation strategy is to smooth the observations to a common, coarser resolution. This allows a direct comparison to theoretical models, while sacrificing only small-scale information. 

In order to explore the dependence of our results on smoothing scale, we consider the filtered RM field, denoted by $\mathrm{RM}_\sigma(\theta)$. In Fourier space, $\mathrm{RM}_\sigma(\ell) = F(\ell) \mathrm{RM}(\ell)$, where $F(\ell)$ describes the Fourier transform of the smoothing filter, which encodes the effect of the finite angular resolution of the observations and any additional filtering applied to the data. 
The smoothed and then squared map can be expressed as:
\begin{align}
A_\sigma(\boldsymbol{\ell})
&= \int \frac{d^2\ell_1}{(2\pi)^2}
\mathrm{RM}_\sigma(\boldsymbol{\ell} - \boldsymbol{\ell}_1)
\mathrm{RM}_\sigma(\boldsymbol{\ell}_1) \nonumber \\
&= \int \frac{d^2\ell_1}{(2\pi)^2}
F(|\boldsymbol{\ell} - \boldsymbol{\ell}_1|)F(\ell_1)
\mathrm{RM}(\boldsymbol{\ell} - \boldsymbol{\ell}_1)
\mathrm{RM}(\boldsymbol{\ell}_1).
\label{eq:A_sigma_conv}
\end{align}
The corresponding analytical expression for the smoothed RM$^2$–halo cross-correlation function is:
\begin{equation}
w_{\mathrm{RM}_\sigma^2,g}(r_\perp=\chi \theta; z) =
2K W_{\mathrm{RM}}^2
\int \frac{dk}{2\pi}   k   J_0\left[k (\chi \theta)\right] P_{e,g}(k; z) \times \int \frac{d^2q}{(2\pi)^2}
F(|\boldsymbol{k} - \boldsymbol{q}|) F(q) P_{\tilde{B}}(q; z).
\label{eq:wrm_f}
\end{equation}

Figure~\ref{fig:filter_wrm} shows the result of the RM-smoothing on the RM$^2$-halo cross-correlation function at $z=0$. Here we employ Gaussian filters with $\sigma = 0.5-2$ pixels, corresponding to comoving scales of $\sigma = 0.21-0.82$ Mpc/$h$. The results are strongly sensitive to the smoothing scale, with $w_{\mathrm{RM}^2_\sigma,\mathrm{g}}(r_\perp;z)$ dropping by about two orders of magnitude in the case of the $\sigma = 2$ pixel smoothing scale. This strong dependence is expected given the steep $q_\perp$ dependence of the $\tilde{B}$ power spectrum. 

For example, consider the case that $P_{\tilde{B}}(q_\parallel=0,q_\perp) = \mathrm{constant}$, i.e., that $\tilde{B}$ has a white-noise power spectrum. A short calculation then shows that $\int d^2q_\perp/(2 \pi)^2 P_{\tilde{B}}(q_\parallel=0,q_\perp)$ drops by a factor of $\frac{\pi}{4\sigma^2 k_{\mathrm{Nyq}}^2}\mathrm{erf}^2(\sigma k_{\mathrm{Nyq}})$ upon Gaussian smoothing, where $k_{\mathrm{Nyq}}$ is the Nyquist frequency. For a two-cell smoothing, this gives a reduction by a factor of $\approx 16 \pi \sim 50$. The actual suppression is even larger, because $P_{\tilde{B}}(q_\parallel=0,q_\perp)$ grows a little more rapidly towards high $q_\perp$ than in the white-noise case. Figure~\ref{fig:filter_wrm}
shows the results of the full calculation following Eq.~\ref{eq:wrm_f}, and confirms that the smoothing scale dependence is captured by the analytic model of Eq.~\ref{eq:wrm_f}.

The main implications of these results are as follows. First, the signal could be larger than in our current models if the $\tilde{B}$ power continues increasing towards smaller scales and the observational smoothing is less than that from our simulation mesh. For example, note that current NVSS measurements have an angular resolution of about $45''$ \citep{taylor_rotation_2009}: this is a factor of a $\sim$ few finer than the effective resolution of our simulation mesh at a typical redshift of $z=0.2$. 
On the other hand, Appendix \ref{app:c} investigates the convergence of our RM$^2$-halo cross-correlation results with simulation resolution (for a fixed mesh scale). There, the results are found to {\em decrease} with increasing resolution, perhaps because magnetic field reversals are better-captured at high resolution. Second, however, the figure shows that the signal may drop steeply if the observations lead to more RM smoothing than in our baseline calculations. Related, one should be cautious in directly comparing RM$^2$ measurements with cosmological MHD simulations, which may have limited resolution. However, our formalism provides an important guide here since we have shown explicitly how $\langle \mathrm{RM}^2 \times \mathrm{g} \rangle$ depends on the smoothing scale. Clearly, accurate comparisons between upcoming measurements and models will require carefully matching to the smoothing scale and pixelization of the observations. Related, any additional filtering steps (e.g. to mitigate galactic foreground contamination) must be carefully included in the model.

\section{Measurement Forecasts}\label{sec:uncertainties}

We now turn to forecast the prospects for measuring the RM$^2$-galaxy cross-correlation in current and upcoming surveys. We account for three main sources of RM measurement noise: i) noise from residual galactic foreground contributions to the RM 
($\sigma_{\mathrm{GRM}}$), ii) scatter induced from intrinsic source contributions to the RM estimates ($\sigma_{\mathrm{in}}$), and iii) instrumental/sky noise in the RM measurements ($\sigma_{\mathrm{obs}}$). In principle, additional RM contributions along the line of sight that are at disparate redshifts from the foreground galaxy sample also contribute to the variance of the cross-correlation estimate. In practice, this makes a sub-dominant contribution to the cross-correlation error budget, and we neglect it in what follows. In Appendix~\ref{app:b} we discuss an explicit test of neglecting sample variance in the $\mathrm{RM}^2$ fields, finding that their neglect is well-justified, at least in the Gaussian fields approximation.
We assume that noise sources i-iii are uncorrelated with the galaxy distribution, and so contribute to the cross-correlation error budget, but do not impact the average $\langle \mathrm{RM}^2 \times \mathrm{g} \rangle$ signal. 

\subsection{RM Noise Budget}

First, let us consider the instrumental/sky part of the error budget, due to thermal measurement noise.  
In order to arrive at plausible numbers here, we briefly summarize the noise estimates from current RM catalogs. The median RM instrumental noise from the NVSS \citep{condon_nrao_1998} catalog, for sources with RM measurements and optical cross-matches, is $\sigma_{\mathrm{obs}} = 10.8~\mathrm{rad \, m^{-2}}$ \citep{hammond_new_2013}. The LOFAR Two-metre Sky Survey (LoTSS) reports a median uncertainty of
$\sigma_{\mathrm{obs}} = 0.06~\mathrm{rad \, m^{-2}}$ \citep{osullivan_faraday_2023}, but the greater precision here is largely due to its operating at lower frequencies. Next, the POSSUM survey, conducted with ASKAP, reports a median uncertainty of
$\sigma_{\mathrm{obs}} = 1.2~\mathrm{rad \, m^{-2}}$ \citep{anderson_probing_2024}. In what follows, we take the POSSUM number to be representative of what can be achieved in future surveys with the SKA \citep{gaensler_origin_2004}, at similar frequencies near $1-1.4$ GHz.  

For the galactic foreground contribution, note that one can use measurements of pulsar RMs in different sky directions to try and estimate and subtract the galactic  RM (GRM) term. The relevant noise term for the GRM then reflects the foreground residuals leftover after subtracting an estimate of the GRM along the line of sight. Here, we adopt a GRM uncertainty of
$\sigma_{\mathrm{GRM}} = 1.2~\mathrm{rad \, m^{-2}}$ based on the POSSUM survey results \citep{anderson_probing_2024}. This should be a conservative estimate as additional improvements may be possible from focusing on high galactic latitude regions, and from using additional pulsar RM measurements (which will help with foreground removal). 

Finally, we consider the intrinsic RM contributions. The intrinsic RM is expected to follow a symmetric zero-mean distribution, as it is equally likely to be positive or negative. Reference~\cite{anderson_probing_2024} estimates an rms intrinsic RM of $\sigma_{\mathrm{in}} = 6.4~\mathrm{rad \, m^{-2}}$ from the POSSUM survey. We adopt that value in what follows.  

The total RM noise, $\sigma_{\mathrm{tot}}$, can then be computed by adding these three contributions in quadrature:
\begin{equation}
    \sigma_{\mathrm{tot}} = \sqrt{ \sigma_{\mathrm{obs}}^2 + \sigma_{\mathrm{GRM}}^2 + \sigma_{\mathrm{in}}^2 }.
 \end{equation}
Adopting the above values from the literature: $\sigma_{\mathrm{obs}} = 1.2~\mathrm{rad \, m^{-2}}$, $\sigma_{\mathrm{GRM}} = 1.2~\mathrm{rad \, m^{-2}}$, and $\sigma_{\mathrm{in}} = 6.4~\mathrm{rad \, m^{-2}}$, gives $\sigma_{\mathrm{tot}} = 6.6~\mathrm{rad \, m^{-2}}$. Hence, the dominant contribution to the noise is from the intrinsic source RM term. 

\subsection{$\langle \mathrm{RM^2 \times g} \rangle$ Forecast Results}

\begin{figure*}[ht!]
    \centering
    \includegraphics[width=0.49\linewidth]{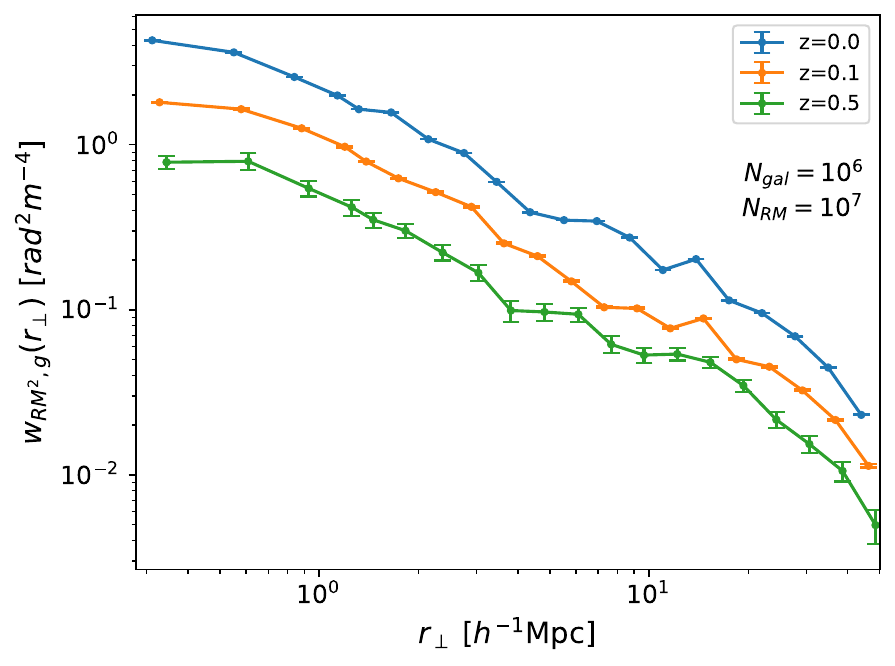} \\
    \includegraphics[width=0.49\linewidth]{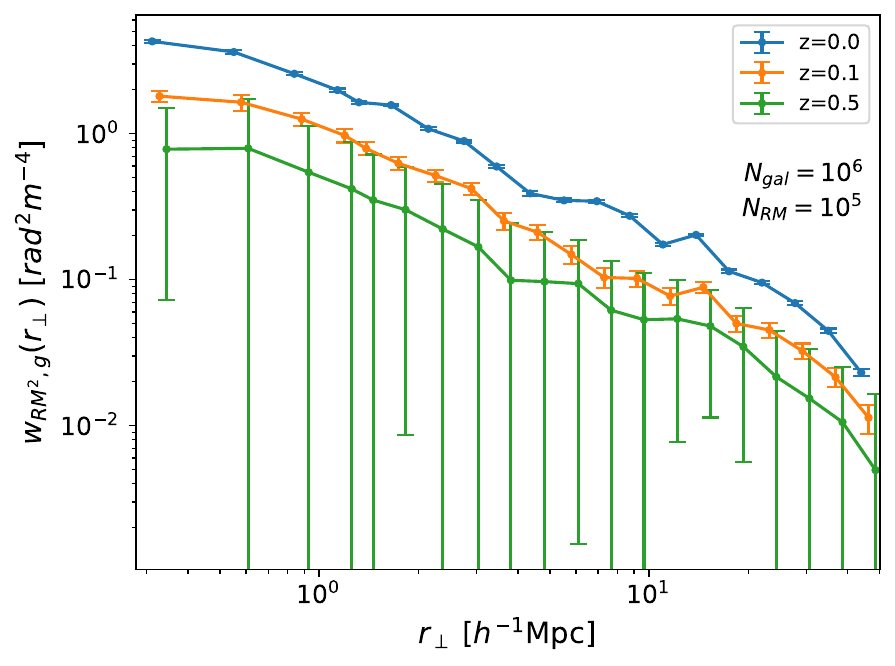}
    \includegraphics[width=0.49\linewidth]{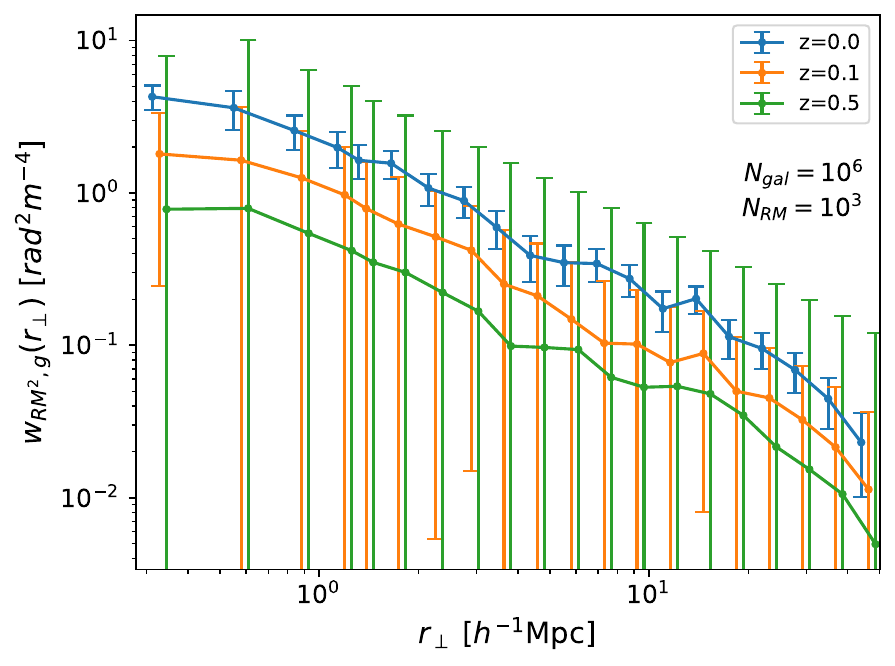}
    \caption{Error bar forecasts for measuring $\langle {\mathrm {RM}}^2 \times {\rm g} \rangle$ in various narrow redshift bins. In each case we assume correlations between RM grids and $10^6$ galaxies, and the error bars are $1-\sigma$ uncertainties. The different panels vary in the RM sample sizes considered with $N_{\mathrm{RM}} = 10^7$ ({\em top row}), $10^5$ ({\em bottom left)}, and $10^3$ ({\em bottom right}). The cross-correlations are extrapolated
    beyond $r_\perp \geq 20$ Mpc/$h$ using the approximate model of Equation~\ref{eq:wrm_approx} to better cover scales where direct simulation estimates are noisy.  The bottom row case is intended to represent current RM samples, the top row is for the expected SKA data set, and the bottom left is an intermediate case. In each scenario, we assume $\sigma_{\rm tot} = 6.6~\mathrm{rad \, m^{-2}}$. The SKA sample promises precise $\langle {\mathrm {RM}}^2 \times {\rm g} \rangle$ measurements across a broad range of spatial scales and redshifts. The intermediate sample also allows high signal-to-noise-ratio measurements, while even current RM catalogs may enable detections for lower redshift bins. 
    For visual clarity, we shift the points and error bars in different redshift bins by $\pm 5\%$, so that the error bars do not overlap, although we adopt the same radial binning at each redshift. 
    }
    \label{fig:error}
\end{figure*}

We can now forecast the expected error bars on $w_{\mathrm{RM^2,g}}(r_\perp;z)$. The forecasts presented here should be interpreted as baseline sensitivity estimates under the Gaussian covariance approximation (see Appendix~\ref{app:b} for details). Additional contributions to the covariance, such as trispectrum terms, could increase the variance. We also work in the
noise-dominated limit, neglecting sample variance contributions to the $\mathrm{RM}^2$ fields (which Appendix~\ref{app:b} argues is well-justified), and in the galaxy overdensity field, as also discussed in Appendix~\ref{app:b}. 
The Gaussian calculation still provides a useful best-case estimate of the noise floor and indicates that the measurement may be feasible by combining current/upcoming RM and galaxy surveys. A full treatment of the non-Gaussian covariance is left for possible future work.

We consider the case of correlations with narrow foreground galaxy redshift distributions, in which the signal follows the results of the previous section. In the Gaussian, and noise-dominated approximation, Appendix~\ref{app:b} shows that the statistical error in each radial bin is given by:
\begin{equation}
\label{eq:wrmsqg_error}
    \sigma_{\text{ring}} = \frac{\sqrt{2} \cdot \sigma_{\mathrm{tot}}^2}{\sqrt{N_{\mathrm{RM}} \cdot N_{\mathrm{galaxy}}}},
\end{equation}
where $N_{\mathrm{RM}}$ is the number of RM sources in the bin, and $N_{\mathrm{galaxy}}$ is the number of galaxies used in the stacking analysis. Note that $N_{\mathrm{RM}}$ depends on the area of the radial bin. We further assume that the error bars in the different radial bins are uncorrelated, which should be a good approximation in the noise-dominated limit. 

Figure~\ref{fig:error} shows the resulting error bar estimates for $w_{\mathrm{RM}^2,~g}(r_\perp;z)$ in several narrow foreground galaxy redshift bins
around $z=0, 0.1$, and $0.5$. These results are shown for three different RM samples: an SKA-like catalog with $10^7$ RMs (top), a sample with $10^5$ RMs (bottom left) -- broadly comparable to the union of all currently available catalogs -- and a case with $10^3$ RMs (bottom right), along the lines of the catalog used in the analysis of reference \cite{amaral_constraints_2021}.  
The only assumption required about the RM redshifts is that they are known to be beyond the galaxy redshift bins. In \cite{amaral_constraints_2021} an optical cross-matched catalog was used to extract only RM sources that are established to be at higher redshift than the tracer galaxies. In practice, a similar cross-matching effort may be required for the larger upcoming RM samples. In each galaxy redshift bin and scenario, we assume that cross-correlation measurements can be carried-out with the entire set of background RM sources. Note that in this sense our $z=0$ results represent a limiting case, as one cannot construct a large foreground galaxy catalog precisely at $z=0$. We include this bin as a useful reference point to show how the forecasts vary with signal strength even though it is not a directly observable configuration.

We consider cross-correlations with $10^6$ galaxies in each redshift bin. This is comparable to the size of the sample already analyzed in \cite{amaral_constraints_2021} from the WISExSuperCOSMOS catalog, which contains over 20 million galaxies with photometric redshifts \citep{bilicki_wise_2016}. Larger samples, across a wider redshift range, are available now from DESI \citep{DESI_sample25}, with further improvements expected soon. The DESI galaxies have spectroscopic redshifts \citep{DESI_sample25}, although even photometric redshifts -- provided they are well-calibrated -- are sufficient for our analysis. 

In the case of the SKA-like survey, the total SNR is enormous and our forecasts suggest that $w_{\mathrm{RM^2,g}}(r_\perp;z)$ can be detected across a wide range of scales from $r_\perp \sim 0.5-50$ Mpc/$h$ at high significance. 
Note that the statistical error bars decrease towards large scales as larger annuli contain more RM$^2$-galaxy pairs. For logarithmic radial bins, the error decreases as $1/\sqrt{N_{\mathrm RM}} \propto 1/r_\perp$. On the other hand, the signal scales roughly as $w_{\mathrm{RM^2,g}}(r_\perp;z) \propto r_\perp^{-0.8}$ (Eq.~\ref{eq:wrm_approx}), so the large radial bins have slightly higher SNR. However, it may be more challenging to detect the weaker signal at large radius because of systematic effects that are neglected in these initial forecasts. 

The redshift evolution of the signal can also be captured, with strong detections expected across all of the redshift bins considered here. This is encouraging for the goal of measuring the redshift evolution of the average cosmic magnetic energy density. 
Although the signal drops steeply with increasing redshift, it is likely possible to make measurements at higher redshifts, provided clean samples of sufficiently high redshift RM sources, and tracer galaxies, can be assembled.  

The bottom two panels also suggest that initial detections may be feasible with current data sets. Here, the scaling is straightforward to understand as these current samples are, respectively, 2 and 4 orders of magnitude smaller than the ultimate SKA ones. Hence, the error bars in these cases are, respectively, 1 and 2 orders of magnitude larger than for the SKA case. As an illustrative example, the total SNR forecast at $z=0.1$, summed over all radial bins is 36.5 for $10^5$ RMs and 3.65 for $10^3$ RMs. This motivates carrying out pilot measurements with the existing data sets. 

\subsection{$\langle |\mathrm{RM}| \times \mathrm{g} \rangle$ and Noise Bias} 
\label{sec:obs_noise}
\begin{figure*}[ht!]
    \centering
    \includegraphics[width=0.49\linewidth]{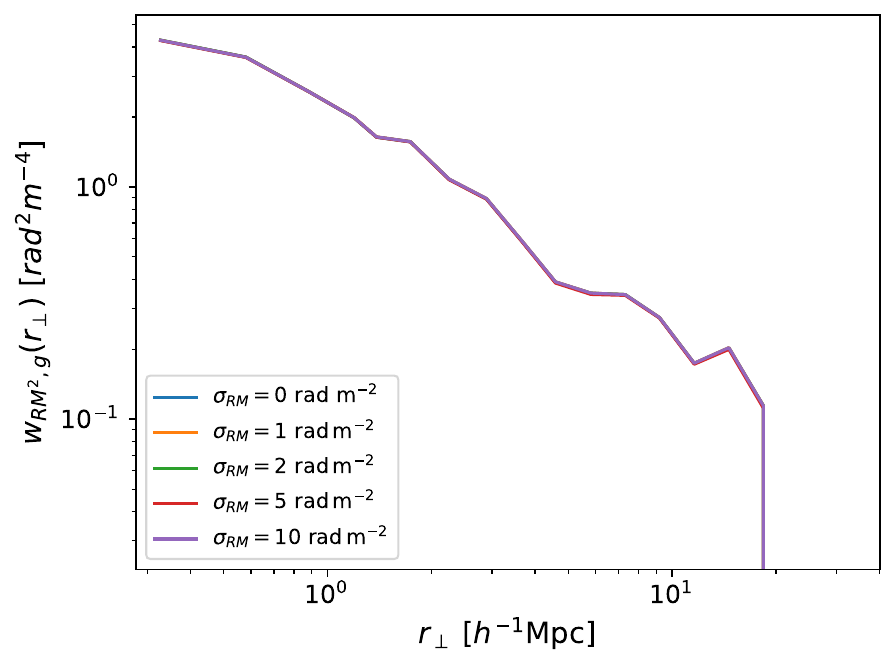}
    \includegraphics[width=0.49\linewidth]{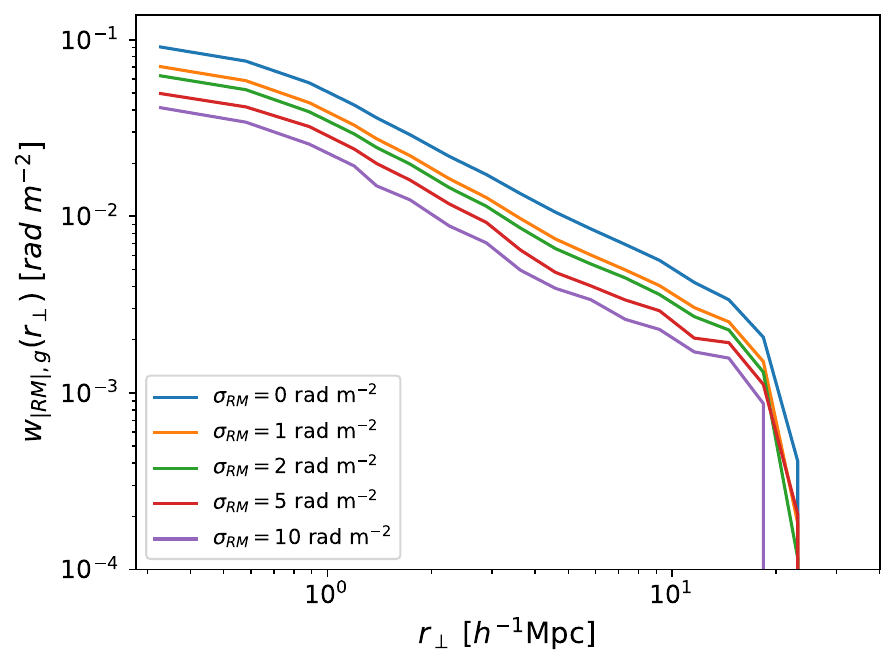}
    \caption{Illustration of the impact of noise bias on two RM-galaxy cross correlation estimators: $w_{\mathrm{RM^2, g}}(r_\perp;z)$ ({\em left panel}) and $w_{|\mathrm{RM}|,{\rm g}}(r_\perp;z)$ ({\em right panel}). In each case, simulated RM grids are degraded by including zero-mean Gaussian noise with standard deviations of
    $\sigma_{\mathrm{RM}} = \{1, 2, 5, 10\}~\mathrm{rad \, m^{-2}}$. These cases span plausible estimates of the noise for current/upcoming surveys (see text), including detector noise, residual foregrounds, and intrinsic source contributions to the effective RM noise. The $w_{|\mathrm{RM}|,{\rm g}}(r_\perp;z)$ estimator is subject to noise bias, as expected, while $w_{\mathrm{RM^2, g}}(r_\perp;z)$ remains robust to noise. In the case of $w_{|\mathrm{RM}|,{\rm g}}(r_\perp;z)$, the noise can strongly bias estimates of the strength of this correlation unless efforts are made to correct for it. Of course, the noise still impacts the variance of $w_{\mathrm{RM^2, g}}(r_\perp;z)$, but it does not suffer from an average bias.}
    \label{fig:noise_bias}
\end{figure*}

We can also consider the $\langle |\mathrm{RM}| \times \mathrm{g} \rangle$ estimator. First, we caution that this
statistic is biased by noise.  After taking the absolute value of the noisy RM field, the expectation value of the signal is modified by the noise: that is, $\langle |\mathrm{RM} + n| \rangle \neq \langle |\mathrm{RM}| \rangle$, even if the noise, $n$, has zero mean. The absolute value operation breaks the symmetry of the noise around zero and leads to a bias. 
This also biases the cross-correlation, $w_{\mathrm{|RM}|,g}(r_\perp;z)$, even when the noise is uncorrelated with the galaxy distribution. The bias arises because the difference between $\langle |\mathrm{RM} + n| \rangle$ and $\langle |\mathrm{RM}| \rangle$ depends on $|\mathrm{RM}|$ itself, which is correlated with galaxy overdensity. To make this more explicit, we derive the expectation value for the noisy |RM| field, $\langle |\mathrm{RM} + n| \rangle_{\mathrm{noise}}$, averaged over realizations of Gaussian noise:
\begin{equation}
    \langle |\mathrm{RM} + n| \rangle_{\mathrm{noise}} = |\mathrm{RM}| \mathrm{erf} \left(\frac{|\mathrm{RM}|}{\sqrt{2}\sigma_n}\right)+\sqrt{\frac{2\sigma_n^2}{\pi}} \exp\left(-\frac{|\mathrm{RM}|^2}{2\sigma_n^2}\right).
\end{equation}
This leads to a noise bias for $\langle \mathrm{|RM| \times g} \rangle$ because $\langle |\mathrm{RM} + n| \rangle_{\mathrm{noise}}$ is a non-linear function of $|\mathrm{RM}|$ and $\sigma_n$,
while $\mathrm{|RM|}$ correlates with the galaxy distribution. Especially in the high-noise regime typical of RM measurements, the estimator is dependent on both the intrinsic signal amplitude $|\mathrm{RM}|$ and the noise $\sigma_n$. In contrast, our $w_{\mathrm{RM^2,g}}(r_\perp;z)$ estimator is unbiased by noise, at least as long as the RM$^2$-noise is uncorrelated with the galaxy distribution, as it is linear in the squared signal. This is straightforward to see:
\begin{align*}
    \left\langle (\mathrm{RM} + n)^2\times  \mathrm{g} \right\rangle 
    &= \left\langle \mathrm{RM}^2 + 2 \mathrm{RM} \, n + n^2 \times  \mathrm{g} \right\rangle \\
    &= \left\langle \mathrm{RM}^2\times  \mathrm{g} \right\rangle 
    + \underbrace{\left\langle 2 \mathrm{RM} \, n\times  \mathrm{g} \right\rangle}_{\mathclap{=~0}} 
    + \underbrace{\left\langle n^2\times \mathrm{g} \right\rangle}_{\mathclap{=~0}} \\
    &= \left\langle \mathrm{RM}^2\times  \mathrm{g} \right\rangle.
\end{align*}
Here, the second term vanishes because the noise $n$ has zero mean and is uncorrelated with both the galaxy distribution and the RM field. The third term vanishes because $n^2$ is uncorrelated with the galaxy overdensity, which has zero mean 
(note that $\mathrm{g}$ here is short-hand for $\delta_\mathrm{g}$). 

In order to quantify how biased $w_{|\mathrm{RM}|,\mathrm{g}}(r_\perp;z)$ is, we add Gaussian random noise with zero mean and standard deviations of $\sigma_{\mathrm{RM}} = \{1, 2, 5, 10\}~\mathrm{rad \, m^{-2}}$ to the mock RM maps.
Figure~\ref{fig:noise_bias} compares the two estimators, $w_{|\mathrm{RM}|,\mathrm{g}}(r_\perp;z)$, and $w_{\mathrm{RM^2,g}}(r_\perp;z)$, in the presence of this noise.  As expected, $w_{\mathrm{RM^2,g}}(r_\perp;z)$ is unbiased by noise.
However, noise leads to a systematic suppression in $w_{|\mathrm{RM}|,\mathrm{g}}(r_\perp;z)$, where the bias depends on the noise level. The suppression is roughly at the factor of $\sim 2$ level for the most realistic cases with $\sigma_{\mathrm{RM}} \sim 5-10 \, \mathrm{rad \, m^{-2}}$. The signal is suppressed here, rather than enhanced, because the noise bias is largest when the true $\mathrm{|RM|}$ is small,
while large $\mathrm{|RM|}$ values occur preferentially in galaxy overdensities. This leads to a negative correlation between the $\mathrm{|RM|}$ noise bias and $\delta_\mathrm{g}$, and a suppression in the cross-correlation.

Note that the previous study of reference \cite{stasyszyn_measuring_2010} considers $|\mathrm{RM}|$ cross-correlations, while normalizing their estimator by $\langle |\mathrm{RM}| \rangle$. This suffers from additional noise bias, since the mean value of $|\mathrm{RM}|$, in the denominator of their estimator, is also biased by noise. In addition, employing the normalized estimator requires knowledge of the average |RM| values -- where the relevant average includes only contributions from structures near the tracer galaxy redshifts -- but these are not known {\em a priori}. 

We can also compare our forecasts with the work of \cite{amaral_constraints_2021}, which used a variant of the
$|\mathrm{RM}|$-galaxy cross-correlation estimator considered here. Their estimator essentially measures the foreground surface density of galaxies around RM sources, weighted by the excess value of $|\mathrm{RM}|$, above that around random locations.
That work cross-correlated 1,742 background RMs from NVSS with the foreground galaxy distributions from WISExSuperCOSMOS.
Translating from their statistic to our related one, $w_{|\mathrm{RM|,g}}(r_\perp;z)$, their results correspond to a $3-\sigma$ upper bound of $3.8~\mathrm{rad\,m^{-2}}$ on scales of $r_\perp = 1-2.5$ Mpc. Our forecasts for this data set yield an SNR of
$0.9$ for the same scales, so the lack of detection in the earlier study is not surprising. Extending their analysis to larger scales (out to $r_\perp \sim 50$ Mpc/$h$), we forecast an SNR of $1.5$, tantalizingly close to a detection. Of course, this is an idealized forecast, and it neglects the noise bias concern for this estimator, among other simplifications.
Matching to the properties of their sample, we forecast an SNR of $3.9$ at $z=0.1$ for the $w_{\mathrm{RM^2,g}}(r_\perp;z)$ statistic. Although the noise is larger for this statistic, the signal is too (see Figure~\ref{fig:rm-gal}), and so the SNR turns out to be a little higher.  

Hence, we conclude that a detection of this signal may be within striking distance. Recall that our model signals are computed by cross-correlating with the entire \textsc{Illustris-TNG} halo catalog: this likely underestimates the large-scale cross-correlation signals at the factor of $\sim 2$ level, because the observed WISE and DESI galaxies tend to reside in larger and more highly clustered host dark matter halos than the typical halos in \textsc{Illustris-TNG}. In addition, high-resolution RM observations in the future may span smaller angular scales than are well-captured in our current models. This could boost the amplitude of the cross-correlation signals relative to our current predictions. Future work should refine our forecasts by matching the models more closely to the expected galaxy properties, further investigating the smoothing-scale dependence, incorporating sample variance in the error budget, and assessing systematic errors from foreground cleaning and other practical concerns. Furthermore, compiling a larger set of background RMs and using DESI galaxy catalogs could enable near-term detections. Despite these caveats, we expect the analytic framework presented in our paper to apply beyond the specifics of the Illustris-TNG simulations.

\section{Conclusion} \label{sec:conclusion}

We presented a new technique to extract more information about cosmic magnetic fields from current and forthcoming Faraday RM data sets. Our method uses cross-correlations between the squared RM field and the galaxy distribution, $w_{\mathrm{RM^2,g}}(r_\perp;z)$. This statistic allows one to cleanly extract contributions to the RM measurements which originate near the redshifts of the foreground tracer galaxies. The RMs intrinsic to the source, from other structures at disparate redshifts, and from the Milky Way impact only the variance of this cross-correlation and not its average value. 

We showed that $w_{\mathrm{RM^2,g}}(r_\perp;z)$ is related to a bispectrum involving two factors of the line-of-sight component of the electron-density--weighted magnetic field strength and one factor of the galaxy density fluctuations. We found a useful
approximation in which the scale-dependence of this statistic is primarily governed by the electron-galaxy cross-correlation function, while the amplitude is mainly determined by an integral over the electron-density--weighted magnetic field strength. This form should apply broadly beyond the specific case of the \textsc{Illustris-TNG} model. 

Using simulated RM maps and halo catalogs from the TNG300-3 simulation, we find that the $\langle \mathrm{RM^2 \times g} \rangle$ signal increases by about three orders of magnitude from $z=3$ to $z=0$. This evolution is driven
by the amplification of magnetic fields in overdense regions by dynamo processes, aided by magnetized plasma transported into the surrounding gas from AGN and galactic outflows. 
The resulting electron-density--weighted magnetic field strength shows a broad distribution with a long tail towards high field strength, and a nearly white-noise spatial power spectrum. Consequently, the $\langle \mathrm{RM^2 \times g} \rangle$ statistic is primarily sensitive to the enhanced field strengths in the inner regions of dark matter halos, and exhibits a significant dependence on the observational RM smoothing scale.  
Nevertheless, it offers a promising means for determining the redshift evolution of electron-weighted magnetic fields.

Our $\langle \mathrm{RM^2 \times g} \rangle$ statistic has advantages over the related statistics considered in previous work, which are variants of $\langle \mathrm{|RM| \times g} \rangle$. First, it connects more directly to the statistical properties of the underlying magnetic field, electron density, and galaxy distribution fluctuations. This makes the modeling and interpretation more transparent. Second, $\langle \mathrm{RM^2 \times g} \rangle$ is immune to noise bias, in contrast to the $\langle \mathrm{|RM| \times g} \rangle$ estimator. Third, our statistic yields a slightly higher SNR. On the other hand, $\langle \mathrm{|RM| \times g} \rangle$ weights highly overdense regions less strongly and may therefore be complementary. Pursuing both estimators jointly will likely be advantageous. 

We forecasted the prospects for measuring $w_{\mathrm{RM^2,g}}(r_\perp;z)$ using current and future RM catalogs and galaxy surveys. In the long term, for an SKA-like survey with $\sim 10^7$ RM measurements over most of the sky, combined with $\sim 10^6$ galaxies per redshift slice, we expect highly significant $w_{\mathrm{RM^2,g}}(r_\perp;z)$ detections across a wide range of spatial scales and redshifts.  Even in the near term, signal detections appear achievable with existing NVSS RM data, particularly when supplemented with recent SKA pathfinder measurements. 

In future studies, it will be important to extend our treatment in several ways. First, we used the entire \textsc{Illustris-TNG} halo catalog, rather than a mock galaxy sample. A proper synthetic galaxy sample is needed for realistic comparisons with observational data.  
Second, it will be important to ensure that the models closely match the smoothing scales of observational results given the sensitive dependence there. Third, our forecasts should be refined to include sample variance and non-Gaussian contributions to the error budget. Next, a more rigorous investigation of the bispectrum of Eq.~\ref{eq:triangle_power} may improve our approximate analytic expression (Eq.~\ref{eq:wrm_approx}).
Finally, constructing detailed mocks will help develop data analysis pipelines, assess systematic effects, and prepare for measurements with upcoming data.

In summary, the $\langle \mathrm{RM^2 \times g} \rangle$ statistic provides a powerful new tool for tracing the evolution and spatial distribution of cosmic magnetic fields. Our simulation studies and analytic modeling should help interpret future measurements. Together with the increasing size and fidelity of RM catalogs and galaxy survey data, this should advance our understanding of the cosmological origin, amplification, and spatial distribution of large-scale cosmic magnetic fields. 


\begin{appendix}
\section{Derivations of $\langle \mathrm{RM^2 \times g} \rangle$ Expressions}
\label{app:a}

Here we give derivations for the key expressions presented in this work, and motivate the approximation analytic formula
we use to interpret the \textsc{Illustris-TNG} simulation results. We also show how to generalize from the main case explored
in the body of this paper, which assumes a narrow galaxy redshift bin, to scenarios with arbitrary galaxy redshift distributions. 

\subsection{Harmonic Space Statistics}
\label{s:ell_space}

We begin with the squared RM field in harmonic space and work in the flat sky approximation. 
We start from Eq.~\ref{eq:a_conv}, which expresses the harmonic space counterpart of the RM-squared field as a convolution in harmonic space. We then combine with Eq.~\ref{eq:proj_gal} for the projected galaxy distribution, and calculate the 
angular cross-power spectrum. The angular cross-power spectrum is the harmonic space counterpart of $\avg{\mathrm{RM}^2 \times \mathrm{g}}$ and is defined by:
\begin{equation}
\label{eq:cl_ag}
C_{\mathrm{A,g}}(\ell) = (2 \pi)^2 \delta_D(\boldsymbol{\ell + \ell^\prime}) \avg{A(\boldsymbol{\ell)} \delta_{\rm g}(\boldsymbol{\ell^\prime})}.
\end{equation}
Inserting Eq.~\ref{eq:a_conv} for $A(\boldsymbol{\ell})$, we can then write:
\begin{equation}
    \label{eq:ag}
    \avg{A(\boldsymbol{\ell})\delta_{\rm g}(\boldsymbol{\ell^\prime})} = \int \frac{d^2\ell_1}{(2 \pi)^2} \avg{\mathrm{RM}(\boldsymbol{\ell - \ell_1}) \mathrm{RM}(\boldsymbol{\ell_1}) \delta_{\rm g}(\boldsymbol{\ell^\prime})}.
\end{equation}

The RM in a given direction is an integral over the electron-density--weighted magnetic field strength. As in the body of the paper, we use the following notoation:
\begin{equation}
\label{eq:mag_den_weighted}
\tilde{B_\parallel}(\boldsymbol{\theta},z) = B_\parallel(\boldsymbol{\theta},z) \left[1 + \delta_e(\boldsymbol{\theta},z)\right],
\end{equation}
i.e., $\tilde{B_\parallel}$ denotes the electron-density--weighted  line-of-sight component of the magnetic field. In analogy to the kSZ case \citep{Dore:2003ex}, note that only components of $\boldsymbol{\tilde{B}}$ which are transverse to the wavevector $\boldsymbol{k}$ will contribute to the projected RM field statistics. 
This is because line-of-sight components of $\boldsymbol{\tilde{B}}$ (or momentum density in the case of kSZ) suffer from cancellations, at least for the relatively small scales of interest here. 
Although 
$\boldsymbol{B}$ is necessarily transverse to $\boldsymbol{k}$, the same is not true of $\boldsymbol{\tilde{B}}$. Explicitly, the transverse
Fourier components of $\boldsymbol{\tilde{B}}$ may be written as: 
\begin{align}
\label{eq:btilde_transverse}
\boldsymbol{\tilde{B}}_\perp(\boldsymbol{k}) &=  \boldsymbol{\tilde{B}}(\boldsymbol{k}) - (\boldsymbol{\hat{k}} \cdot \boldsymbol{\tilde{B}}) \boldsymbol{\hat{k}} \nonumber \\
& = \boldsymbol{B}(\boldsymbol{k}) + \int \frac{d^3k^\prime}{(2 \pi)^3} \delta_e(\boldsymbol{k-k^\prime}) \left[\boldsymbol{B}(\boldsymbol{k^\prime}) - \left(\boldsymbol{\hat{k}} \cdot \boldsymbol{B}(\boldsymbol{k^\prime})\right) \boldsymbol{\hat{k}} \right],
\end{align}
where we have made use of the fact that $\boldsymbol{B}$ is purely transverse, and so $\boldsymbol{B}_\perp(\boldsymbol{k})=\boldsymbol{B}(\boldsymbol{k})$.

Another way to see that $\boldsymbol{\tilde{B}}$ is not generally transverse is to consider the real space divergence of this field:
\begin{equation}
\boldsymbol{\nabla} \cdot \boldsymbol{\tilde{B}} = \boldsymbol{\nabla} \cdot \left[(1+\delta_e) \boldsymbol{B}\right] = \boldsymbol{B} \cdot \boldsymbol{\nabla}\left(\delta_e\right) + \delta_e \boldsymbol{\nabla} \cdot \boldsymbol{B}= \boldsymbol{B} \cdot \boldsymbol{\nabla}\left(\delta_e\right).
\end{equation}
This implies that $\boldsymbol{\tilde{B}}$ is only purely transverse in locations where gradients in the electron density distribution vanish or are perpendicular to the local magnetic field direction. 

The line of sight component of the $\boldsymbol{\tilde{B}}_\perp(\boldsymbol{k})$ follows by taking the dot product between Equation \ref{eq:btilde_transverse} and $\boldsymbol{\hat{z}}$, a unit vector in the line-of-sight direction. (Recall that we are working in the small angle approximation, where the line-of-sight direction may be taken to coincide with the $\boldsymbol{\hat{z}}$ direction.) This gives:
\begin{equation}
\boldsymbol{\tilde{B}}_\perp(\boldsymbol{k}) \cdot \boldsymbol{\hat{z}} = \tilde{B}_{\perp,\parallel}(\boldsymbol{k}) =
B_\parallel(\boldsymbol{k}) + \int \frac{d^3k^\prime}{(2 \pi)^3} \delta_e(\boldsymbol{k - k^\prime})  \left[B_\parallel(\boldsymbol{k^\prime}) - \left(\boldsymbol{\hat{k}} \cdot \boldsymbol{B}(\boldsymbol{k^\prime})\right) \left(\boldsymbol{\hat{k}} \cdot \boldsymbol{\hat{z}}\right) \right],
\end{equation}
where note that here the $\perp$ symbol refers to vector components which are perpendicular to the direction of the wavevector $\boldsymbol{k}$. Thus, $\tilde{B}_{\perp,\parallel}(\boldsymbol{k})$ refers to components of $\boldsymbol{\tilde{B}}$ that are perpendicular to the wavevector, yet parallel to the line-of-sight direction. 

Adopting the Limber approximation, we can immediately evaluate the correlation of interest:
\begin{equation}
\label{eq:bisp_projected}
\begin{aligned}
\left\langle A(\boldsymbol{\ell})\,\delta_{\rm g}(\boldsymbol{\ell}') \right\rangle 
&= (2\pi)^2 \delta_D(\boldsymbol{\ell}+\boldsymbol{\ell}')
   \, W_{\rm RM}^2
   \int \frac{d\chi}{\chi^4}\,
   \Theta(\chi_s-\chi)\, W_{\rm g}(\chi)                      \\
&\quad\times
   \int \frac{d^2\boldsymbol{\ell}_1}{(2\pi)^2}\,
   \mathcal{B}_{\tilde{B}_{\perp,\parallel};
               \tilde{B}_{\perp,\parallel};g}
   \!\left(
      \frac{\boldsymbol{\ell}-\boldsymbol{\ell}_1}{\chi},
      \frac{\boldsymbol{\ell}_1}{\chi},
      \frac{-\boldsymbol{\ell}}{\chi}
    \right).
\end{aligned}
\end{equation}
where $\mathcal{B}_{\tilde{B}_{\perp,\parallel}; \tilde{B}_{\perp,\parallel}; g}$ is a bispectrum involving two copies of the $z$-components of the electron-density--weighted magnetic field and the galaxy fluctuation field. Here $W^2_{\mathrm{RM}}$ has factored out of the integral because the weighting factor as defined in Equation~\ref{eq:rm_chi} is redshift independent. 
This result is directly analogous to that of the $\langle \mathrm{kSZ}^2 \times \delta_{\rm g} \rangle$ cross-power spectrum, which can be expressed as an integral over a momentum density-galaxy bispectrum \citep{Dore:2003ex,Ferraro:2016ymw}.  

As in the case of the kSZ effect, this can alternatively be written as an integral over the ``triangle power spectrum'' \citep{Dore:2003ex}:
\begin{equation}
\label{eq:triangle_limber}
C_{\mathrm{A,g}}(\ell) = W^2_{\mathrm{RM}} \int \frac{d\chi}{\chi^4} \Theta(\chi_s-\chi) W_{\rm g}(\chi) \mathcal{T}\left(k = \frac{\ell}{\chi}; z\right),
\end{equation}
with
\begin{equation}
\label{eq:triangle_def}
\mathcal{T} \left(k = \frac{\ell}{\chi}; z\right) = \int \frac{d^2q}{(2 \pi)^2} \mathcal{B}_{\tilde{B}_{\perp,\parallel}; \tilde{B}_{\perp,\parallel}; g} \left(\boldsymbol{k-q}, \boldsymbol{q}, \boldsymbol{-k}; z \right).
\end{equation}
That is, the triangle power spectrum is a two-dimensional integral over the bispectrum of interest. 

\subsection{Real Space Statistics}
\label{s:theta_space}

In this study, we work with real space statistics rather than in harmonic space. Here we discuss some of the real-space counterparts to the harmonic space statistics in the previous subsection. We can determine the angular correlation function between $\delta_{\rm g}$ and $A = \mathrm{RM^2}$ by Fourier transforming Equation~\ref{eq:triangle_limber} yielding the usual, well-known result:
\begin{equation}
    \label{eq:w_ag}
    w_{\mathrm{RM^2,g}}(\theta) = \int \frac{d^2\ell}{(2 \pi)^2} C_{\mathrm{A,g}}(\ell) e^{i \boldsymbol{\ell} \cdot \theta} = \int \frac{d \ell}{2 \pi} \ell J_0(\ell \theta) C_{\mathrm{A,g}}(\ell).
\end{equation}
We can further employ Equation~\ref{eq:triangle_limber} to express this as an integral over the triangle power spectrum:
\begin{equation}
\label{eq:wag_triangle_appendix}
w_{\mathrm{RM^2,g}}(\theta) = W^2_{\mathrm{RM}} \int \frac{d \chi}{\chi^2} \Theta(\chi_s-\chi) W_{\rm g}(\chi) \int \frac{dk}{2 \pi} k J_0(k \chi \theta) \mathcal{T}\left(k;z\right)
\end{equation}
Finally, the integral over $\mathcal{T}\left(k;z\right)$ can also be expressed as an integral over the 3D correlation function between
$\tilde{B_\parallel}^2$ and the three-dimensional galaxy fluctuation field $\delta_{\rm g}$ as:
\begin{align}
\label{eq:xcorr_threed_relation}
w_{\mathrm{RM^2,g}}(\theta) & = W^2_{\mathrm{RM}} \int \frac{d \chi}{\chi^2} \Theta(\chi_s-\chi) W_{\rm g}(\chi) \int \frac{dk}{2 \pi} k J_0(k \chi \theta) \mathcal{T}\left(k;z\right) \nonumber \\
& = \int d\chi \Theta(\chi_s-\chi) W_{\rm g}(\chi) W^2_{\mathrm{RM}} \int_{-\infty}^{\infty} d x_\parallel \xi_{\tilde{B_\parallel}^2, g}\left(\sqrt{x_\parallel^2 + \chi^2 \theta^2}; z\right).
\end{align}
Note that the quantity $\int_{-\infty}^{\infty} d x_\parallel \, \xi_{\tilde{B_\parallel}^2, g}\left(\sqrt{x_\parallel^2 + \chi^2 \theta^2}; z\right)$ has units of $\mu G^2 \, {\rm Mpc}^2$.  

The integrated correlation function term above, multiplied by the factor of $W^2_{\mathrm{RM}}$ can be estimated from coeval simulation boxes. 
This shows how one can estimate $\avg{\mathrm{RM^2} \times {\rm g}}$ for an arbitrary galaxy redshift distribution by stitching together the results from a set of coeval simulation boxes. Explicitly,
\begin{equation}
w_{\mathrm{RM^2,g}}(\theta) = \int d\chi \Theta(\chi_s-\chi) W_{\rm g}(\chi) w_{\rm coeval}(\chi \theta;z),
\end{equation}
with
\begin{equation}
     w_{\rm coeval}(\chi \theta;z) = W^2_{\mathrm{RM}} \int_{-\infty}^{\infty} d x_\parallel \, \xi_{\tilde{B_\parallel}^2, g}\left(\sqrt{x_\parallel^2 + \chi^2 \theta^2}; z\right).
\end{equation}
Here, we take a weighted average of projected correlation functions at various redshifts, while scaling between transverse co-moving coordinate, $r_\perp$, and angle using $\theta = r_\perp/\chi(z)$. This approach avoids constructing full light cones and exploits the Limber approximation, which in real space tells us that the angular correlation functions pick up most of their contributions from pairs of points at the same radial distance. 

\subsection{Approximate Expressions}

So far, the expressions given assume only the flat sky and Limber approximations. The triangle power spectrum of Equation~\ref{eq:triangle_def} and the underlying bispectrum are complex and their full investigation is deferred to possible future work. Here we attempt to determine useful approximations for the bispectrum and the $\avg{\mathrm{RM^2} \times {\rm g}}$ correlation function. We will be satisfied with understanding some of the main qualitative features of these statistics, rather than deriving highly accurate perturbative expressions. 

In this context, it is useful to define the magnetic field power spectrum via:
\begin{equation}
\label{eq:magpower_def}
\avg{B_i(\boldsymbol{k}) B^\star_j(\boldsymbol{k^\prime})} = (2 \pi)^3 \delta_D(\boldsymbol{k} - \boldsymbol{k^\prime}) \left[\delta_{ij} - \hat{k_i}\hat{k_j}\right] P_B(k)
\end{equation}
The projection operator,
\begin{equation}
\label{eq:proj_op}
P_{ij} = \left[\delta_{ij} - \hat{k_i}\hat{k_j}\right],
\end{equation}
appearing in this equation enforces that the magnetic field is transverse. Note that with this definition of the magnetic field power spectrum
the mean-squared magnetic field strength at a point is related to $P_B(k)$ via
\begin{equation}
\avg{\boldsymbol{B(x) \cdot B(x)}} = \int \frac{d^3k}{(2 \pi)^3} 2 P_B(k).
\label{eq:bsquared}
\end{equation}
In some works (e.g. \cite{Kim:1994zh,Adi:2023doe}), there is an additional factor of $1/2$ in front of Equation~\ref{eq:magpower_def}, so that the mean-squared field is just an integral over $P_B(k)$ without the factor of $2$ in Equation~\ref{eq:bsquared} under our convention. 

Returning now to our main objective, we would like to arrive at an approximate expression for the following bispectrum
\begin{equation}
\label{eq:bspec_desired}
   \avg{\tilde{B}_{\perp,\parallel}(\boldsymbol{k_1}),\tilde{B}_{\perp,\parallel}(\boldsymbol{k_2}) \delta_{\rm g}(\boldsymbol{k_3})} 
   = (2 \pi)^3 \delta_D(\boldsymbol{k_1 + k_2 + k_3}) \mathcal{B}_{\tilde{B}_{\perp,\parallel}; \tilde{B}_{\perp,\parallel}; g}(\boldsymbol{k_1},\boldsymbol{k_2},\boldsymbol{k_3})   
\end{equation}
The electron-density-weighted magnetic field strength in Fourier space involves a convolution between the magnetic field, $B_\parallel(\boldsymbol{k})$ and the electron
distribution, $\delta_e(\boldsymbol{k})$. Note that this bispectrum involves five fields, since each real space $\tilde{B_\parallel}$ involves a product 
of the $\boldsymbol{\hat{z}}$-component of the magnetic field and the electron density fluctuation. 
We will start with two strong assumptions to make headway, and then partly relax them in what follows.
First, we will assume the Gaussian field approximation in which the higher-point correlations involved in the above statistic can be factorized into products of two-point correlation functions. Second, we will initially neglect terms involving correlations between $B_\parallel$ and each of $\delta_e$ and $\delta_{\rm g}$. Under these approximations, we can apply Equation~\ref{eq:btilde_transverse} to write

\begin{equation}
\begin{aligned}
 \avg{\tilde{B}_{\perp,\parallel}(\boldsymbol{k_1}),\tilde{B}_{\perp,\parallel}(\boldsymbol{k_2}) \delta_{\rm g}(\boldsymbol{k_3})} 
& \approx \int \frac{d^3k^\prime}{(2 \pi)^3} \avg{\delta_e(\boldsymbol{k_1 - k^\prime}) \delta_{\rm g}(\boldsymbol{k_3})} \avg{B_\parallel(\boldsymbol{k^\prime}) B_\parallel(\boldsymbol{k_2})} + \left(\boldsymbol{k_1} \leftrightarrow \boldsymbol{k_2}\right) \nonumber \\
& = (2\pi)^3 \delta_D(\boldsymbol{k_1 + k_2 + k_3}) \Biggl\{ \left[1 - \left(\boldsymbol{\hat{k}_2 \cdot \hat{z}}\right)^2\right] P_B(k_2) P_{e,g}(k_3) \\
& \hspace{1em} + \left(\boldsymbol{k_1} \leftrightarrow \boldsymbol{k_2}\right) \Biggr\},
\end{aligned}
\end{equation}
where we have used~Equation~\ref{eq:magpower_def} and carried out the integration over $d^3k^\prime$. Here $P_{e,g}(k)$ is the electron density fluctuation-galaxy cross-power spectrum, which can be evaluated from the simulations.  

We can then plug this result into
the triangle power spectrum formula, Equation~\ref{eq:triangle_def}, which yields a simple result:
\begin{equation}
\label{eq:tofk_approx}
\mathcal{T_{\rm rough}}(k) \approx 2 P_{e,g}(k) \int \frac{d^2q}{(2 \pi)^2} P_B(q).
\end{equation}
The factor of two arises from the symmetry above under interchanging $\boldsymbol{k_1} \leftrightarrow \boldsymbol{k_2}$.
Note that the triangle power spectrum involves an integral over transverse modes, for which the $\left[1 - \left(\boldsymbol{\hat{q} \cdot \hat{z}}\right)^2\right]$ factor above is just unity. The quantity $\int \frac{d^2q}{(2 \pi)^2} P_B(q)$ has units of $(\mu G)^2 \, {\rm Mpc}$ and
is closely related to the average of $\boldsymbol{\hat{z}}$-component of the magnetic field, after projecting along the line-of-sight and {\em then} squaring the result.\footnote{Note that $P_B(q)$ itself has units of $(\mu G)^2 \, {\rm Mpc}^3$.} 

A defect of this approximation is that the magnetic field strength likely has strong correlations with the electron density distribution, as seen in the \textsc{Illustris-TNG} simulations (e.g. Figure~\ref{fig:los}). Here we suppose that, for present purposes, the main impact of these is to replace $P_B(q)$ above with 
$P_{\tilde{B}}(q)$. That is, we suggest replacing the magnetic field power spectrum in this formula with the power spectrum of the electron-density--weighted magnetic field strength. 
Although a detailed investigation is beyond the scope of our current study, this replacement may have rigorous motivation from expanding the correlations in Equation \ref{eq:bspec_desired} to higher order in perturbation theory, beyond the simple Gaussian fields assumption for each of $B_\parallel$, $\delta_e$, and $\delta_{\rm g}$. The approximate form, with the $P_{\tilde{B}}(q)$ replacement, allows for the fact that magnetic field strengths are highly enhanced in regions with large upward fluctuations in the electron density field. Further, we allow for an additional redshift-independent calibration factor $K$ which we will match to results from the \textsc{Illustris-TNG} simulations:
\begin{equation}
\mathcal{T}(k) \approx 2 K P_{e,g}(k) \int \frac{d^2q_\perp}{(2 \pi)^2} P_{\tilde{B}}(q_\parallel=0,q_\perp).
\end{equation}
$K$ is essentially a fudge factor to improve agreement with the simulations given that 
replacing $P_B$ with $P_{\tilde{B}}$ lacks rigorous justification. 
Here we have made it explicit that the $\tilde{B}$ power spectrum is evaluated for transverse modes, which was left implicit
earlier. 
We can then plug this result into Equation~\ref{eq:wag_triangle_appendix}, finding:
\begin{equation}
\label{eq:wag_refine}
\begin{aligned}
w_{\mathrm{RM^2,g}}(\theta)
&= 2K\,W_{\mathrm{RM}}^2
   \int d\chi\,\Theta(\chi_s-\chi)\,W_{\rm g}(\chi) \\
&\quad\times
   \left[ \int \frac{d^2 q_\perp}{(2\pi)^2}\,
               P_{\tilde{B}}(q_\parallel=0,q_\perp;z) \right]
   \left[ \int \frac{dk}{2\pi}\, k\, J_0\!\big(k\chi\theta\big)\, P_{e,g}(k;z) \right].
\end{aligned}
\end{equation}
In the narrow galaxy redshift distribution limit, we arrive at Eq.~\ref{eq:wrm_approx} in the body of the text. 
That is, $w_{\mathrm{RM^2,g}}$ is proportional to the product of the projected power of the electron-density-–weighted magnetic field and the electron–galaxy correlation.

In summary, the expressions derived here suggest that the overall scale dependence of $\langle \mathrm{RM}^2 \times \mathrm{g} \rangle$ follows the electron density-galaxy two-point cross-correlation, while the amplitude is set by $\int \frac{d^2q_\perp}{(2 \pi)^2} P_{\tilde{B}}(q_\parallel=0,q_\perp;z)$, the electron-density--weighted magnetic field strength. As discussed in the body of the paper, the correlation between the $\mathrm{RM}^2$ field and the projected galaxy distribution across multiple redshift bins can then be used to extract how the electron-density--weighted magnetic field strength evolves with redshift.

\section{Cross-Correlation Function Covariance Matrix}
\label{app:b}
Here we derive the error covariance matrix for the angular correlation function $w_{\mathrm{RM^2,g}}(\theta)$ and give a useful approximation to this quantity in the noise-dominated limit, as will be defined precisely below. These calculations justify the error bar estimates in our forecasts (\S \ref{sec:obs_noise}) and clarify their regime of validity. 

Our general aim is to calculate the covariance between estimates of the angular correlation function in two narrow angular bins
centered around $\theta_i$ and $\theta_j$, respectively. Using Equation~\ref{eq:w_ag}, this may be written as \citep{Cohn:2005ex}:
\begin{align}
\label{eq:covariance}
\mathrm{Cov}\!\left[w_{\mathrm{RM^2,g}}(\theta_i),\, w_{\mathrm{RM^2,g}}(\theta_j)\right]
&= \frac{1}{(\Delta\theta)^2}
\int_{\theta_i - \Delta\theta/2}^{\theta_i + \Delta\theta/2} d\theta
\int_{\theta_j - \Delta\theta/2}^{\theta_j + \Delta\theta/2} d\theta'
\int \frac{d\ell}{2\pi}\, \ell 
\int \frac{d\ell'}{2\pi}\, \ell'\,
J_0(\ell\theta)\, J_0(\ell'\theta')  \nonumber \\[0.5em]
&\hspace{4em}\times
\Big[
    \langle C_{\mathrm{A,g}}(\ell)\, C_{\mathrm{A,g}}(\ell') \rangle
    - \langle C_{\mathrm{A,g}}(\ell)\rangle\, \langle C_{\mathrm{A,g}}(\ell')\rangle
\Big],
\end{align}
where the integrals over $\theta$ and $\theta^\prime$ describe bin-averaging, $C_{\mathrm{A,g}} (\ell)$ is the angular cross-power spectrum between the $\mathrm{RM}^2$ and $\delta_{\rm g}$ fluctuations, and the $J_0$'s are Bessel functions, as usual. 

Under the approximation of Gaussian statistics, the quantity in square brackets above is given by:
\begin{equation}
\label{eq:gauss_cell_cov}
\mathrm{Cov}\!\left[C_{\mathrm{A,g}}(\ell),\, C_{\mathrm{A,g}}(\ell')\right]
= \frac{2\pi}{\Omega_{\rm sky}}\, \frac{1}{\ell} \,\delta_D(\ell-\ell')\,
\left[
    C_{\mathrm{A,g}}^2(\ell)
    + \left(C_{\mathrm{A,A}}(\ell) + \mathcal{N}_{\mathrm{A,A}}\right)
      \left(C_{\mathrm{g,g}}(\ell) + \frac{1}{n_{\mathrm g}}\right)
\right].
\end{equation}
Note that the Gaussian approximation to the covariance matrix neglects a trispectrum term \citep{Meiksin99,Scoccimarro99}. Here, $C_{\mathrm{A,A}}(\ell)$ denotes the power spectrum of the $\mathrm{A}=\mathrm{RM}^2$ field, including contributions from magnetized plasma at all redshifts, i.e., this term accounts for magnetized plasma beyond the redshift range covered by the galaxy survey. Put differently, portions of the $\mathrm{RM}^2$ field that are produced at redshifts outside of those spanned by the galaxy sample do not contribute to the average $\langle \mathrm{RM}^2 \times \mathrm{g} \rangle$ signal, but they do impact the error bar on this statistic. The quantity $\mathcal{N}_{\mathrm{A,A}}$ describes other noise contributions to the $\mathrm{RM}^2$ signal, which here are assumed to arise from: instrumental noise, intrinsic contributions to the RM from the source itself, and any residual (un-subtracted) portions of the RM from the Milky Way, as discussed quantitatively in \S \ref{sec:obs_noise}. 
While $C_{\mathrm{A,A}}(\ell)$ should trace the large-scale structure, we suppose that $\mathcal{N}_{\mathrm{A,A}}$ is a white-noise field (i.e., with an $\ell$-independent angular power spectrum). In this case, we can write
\begin{equation}
\label{eq:noise_power}
    \mathcal{N}_{\mathrm{A,A}} = 2 \sigma^4_{\mathrm{tot}} \Omega_{\rm pix},
\end{equation}
where $\sigma_{\rm tot}$ is the rms noise in the $\mathrm{RM}$ measurements, while $\Omega_{\rm pix}$ denotes the pixel size of the extragalactic RM map. 
The $2 \sigma^4_{\mathrm{tot}}$ number gives the pixel-to-pixel variance of the $\mathrm{RM}^2$ noise field under Gaussian statistics. Finally, $C_{\mathrm{g,g}}(\ell)$ denotes the galaxy auto-power spectrum, and $1/n_{\mathrm g}$ is the shot-noise in the galaxy distribution, with $n_{\mathrm g}$ denoting the number of galaxies per steradian in the survey. As usual, $C_{\mathrm{A,g}}(\ell)$ is the angular cross-power of the $\mathrm{A}$ and $\delta_{\rm g}$ fields. 

In this work, we take the noise-dominated limit of Equation~\ref{eq:gauss_cell_cov}. In this approximation, we suppose that $\mathcal{N}_{\mathrm{A,A}} \gg C_{\mathrm{A,A}}(\ell)$ and that $1/n_{\mathrm g} \gg C_{\mathrm{g,g}}(\ell)$ at all scales of interest. We further assume that $C_{\mathrm{A,g}}(\ell) \ll \mathcal{N}_{\mathrm{A,A}}/n_{\mathrm g}$ for all $\ell$. These approximations simplify the calculations considerably and are likely well-justified, mainly because the intrinsic source RM generally dominates that from the intervening large-scale structure. 

As a cross-check, we perform an explicit test of neglecting the $\mathrm{RM}^2$ signal power spectrum in the error budget. Specifically, we calculate $C_{\mathrm{A,A}}(\ell)$ in the Gaussian fields approximation, using a model for the $\mathrm{RM}$ angular power spectrum
that approximately matches our simulation results.  We find that the signal from intervening magnetized plasma, for an RM source at $z_s = 1$, is two to three orders of magnitude smaller than the intrinsic noise power, $\mathcal{N}_{\mathrm{A,A}}(\ell)$, across all relevant multipoles, even for SKA-level noise. Physically, this reflects the fact that the variance of the RM field is dominated by intrinsic source contributions rather than intervening plasma fluctuations. 
For this same reason, the cross-term $C_{\mathrm{A},\mathrm{g}}$ in Equation~\ref{eq:gauss_cell_cov} can be safely neglected, as the galaxy-correlated $\mathrm{RM}^2$ signal from the intervening material is small relative to the intrinsic source $\mathrm{RM}^2$ noise. Although we neglect a trispectrum contribution to $C_{\mathrm{A,A}}(\ell)$, including it is unlikely to impact our conclusion given that the Gaussian estimate is so far below the $\mathcal{N}_{\mathrm{A,A}}(\ell)$ noise term.

Taking the shot-noise dominated limit for the galaxy survey is a less accurate approximation at high galaxy number densities. We nevertheless adopt it here for simplicity; see also the discussion below. 

Assuming the noise-dominated limit in Equations~\ref{eq:covariance}  and \ref{eq:gauss_cell_cov}, we arrive at:
\begin{equation}
\label{eq:cov_noiselim}
 \mathrm{Cov}\!\left[w_{\mathrm{RM^2,g}}(\theta_i),\, w_{\mathrm{RM^2,g}}(\theta_j)\right] =
\frac{\mathcal{N}_{\mathrm{A,A}}}{n_{\mathrm g} \Omega_{\rm sky}} \frac{1}{\left(\Delta \theta\right)^2} \int_{\theta_i - \frac{\Delta \theta}{2}}^{\theta_i + \frac{\Delta \theta}{2}} d\theta \int_{\theta_j - \frac{\Delta \theta}{2}}^{\theta_j + \frac{\Delta \theta}{2}} d\theta^\prime \int \frac{d\ell}{2\pi} \ell J_0(\ell \theta) J_0(\ell \theta^\prime).
\end{equation}

This can be simplified by noting that:
\begin{equation}
\label{eq:J0_deltafunc}
\int \frac{d\ell}{2\pi} \ell J_0(\ell \theta) J_0(\ell \theta^\prime) = \frac{\delta_D(\theta - \theta^\prime)}{2 \pi \theta}.
\end{equation}

One can then carry out the bin-averaging integrals -- assuming narrow bins -- and arrive at a simple approximation for the covariance matrix of interest:
\begin{equation}
\label{eq:cov_final_approx}
\mathrm{Cov}\!\left[w_{\mathrm{RM^2,g}}(\theta_i),\, w_{\mathrm{RM^2,g}}(\theta_j)\right]
= \delta_{i j} \frac{2 \sigma^4_{\mathrm{tot}} \Omega_{\rm pix}} {n_{\mathrm g} \Omega_{\rm sky} (2 \pi \theta_i \Delta \theta)},
\end{equation}
after inserting Equation~\ref{eq:noise_power} into the previous expression. 
This equation is easy to interpret: first, in the noise-dominated limit the covariance matrix is diagonal, as expected. Next, note that the quantity $n_{\mathrm g} \Omega_{\rm sky}$ is the number of galaxies in the survey, while $\Omega_{\rm pix}/(2 \pi \theta_i \Delta \theta)$ gives the number of RM measurements contributing to a $w_{\mathrm{RM^2,g}}$ estimate in an annulus of angular area $2 \pi \theta_i \Delta \theta$. Hence, the correlation function variance scales with the variance per pixel,
$2 \sigma^4_{\mathrm{tot}}$, and inversely with the number of measurement pairs in a bin, $n_{\mathrm g} \Omega_{\rm sky} 2 \pi \theta_i \Delta \theta/\Omega_{\rm pix}$.
While this formula is perhaps intuitively obvious, the above analysis clarifies its regime of validity. 
A similar formula holds for the projected correlation function, expressed as a function of transverse comoving distance, in the case of a narrow galaxy redshift bin. This is the main forecasting equation used in our paper (Eq.~\ref{eq:wrmsqg_error}), which follows from the above analysis after swapping the angular bins above for transverse radial bins. 

The biggest shortcoming in our error estimate is likely the neglect of sample variance in the galaxy fluctuations. In representative redshift bins with $0.1 < z < 0.2$, $0.2 < z < 0.3$, and $0.3 < z < 0.5$, we find that the shot-noise power spectrum exceeds the galaxy clustering power for multipoles of roughly $\ell \gtrsim 100$. The corresponding transverse comoving scales in real space are roughly $r_\perp \sim \pi \chi(z)/\ell = 9,  19, 39$ Mpc/$h$, with shot-noise dominating on smaller scales. Hence, neglecting the clustering sample variance contribution should be a good first approximation. 
 Future refinements to our forecasts should include sample variance contributions to the error budget, while also accounting for the non-Gaussian trispectrum term.  Our simplified Gaussian forecasts still provide useful baseline detectability estimates. The high signal-to-noise-ratios expected for the SKA measurements are encouraging and motivate further efforts to explore their practical implementation.

\section{Dependence on Simulation Resolution}
\label{app:c}

\begin{figure}[ht]
    \centering
    \includegraphics[width=0.5\linewidth]{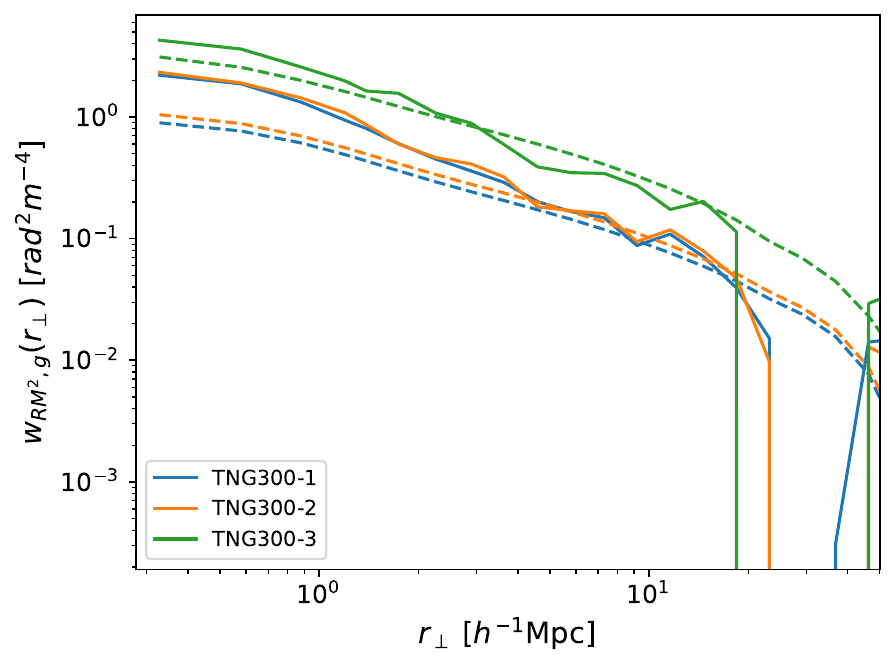}
    \caption{Comparison between the simulated RM$^2$–halo cross-correlation function and the analytical approximation, evaluated at redshift $z = 0$ across three \textsc{Illustris-TNG} resolution levels: TNG300-1, TNG300-2, and TNG300-3. These simulations track $2 \times 2500^3$, $2 \times 1250^3$,
    and $2 \times 625^3$ DM particles plus gas cells, respectively. 
     The RM$^2$–halo correlation amplitude decreases with increasing resolution. This trend is attributed to magnetic field reversals being better resolved in higher resolution simulations, which reduce the net line-of-sight electron-density--weighted magnetic field due to cancellation effects. In lower-resolution runs, directional coherence in under-resolved fields enhances the projected signal.
     The dashed lines show that the trend is fairly well reproduced in the analytic approximation, although the agreement at small radius is less good at higher resolution. }
    \label{fig:rm_sq_res}
\end{figure}

\begin{figure}[ht]
    \centering
    \includegraphics[width=0.6\linewidth]{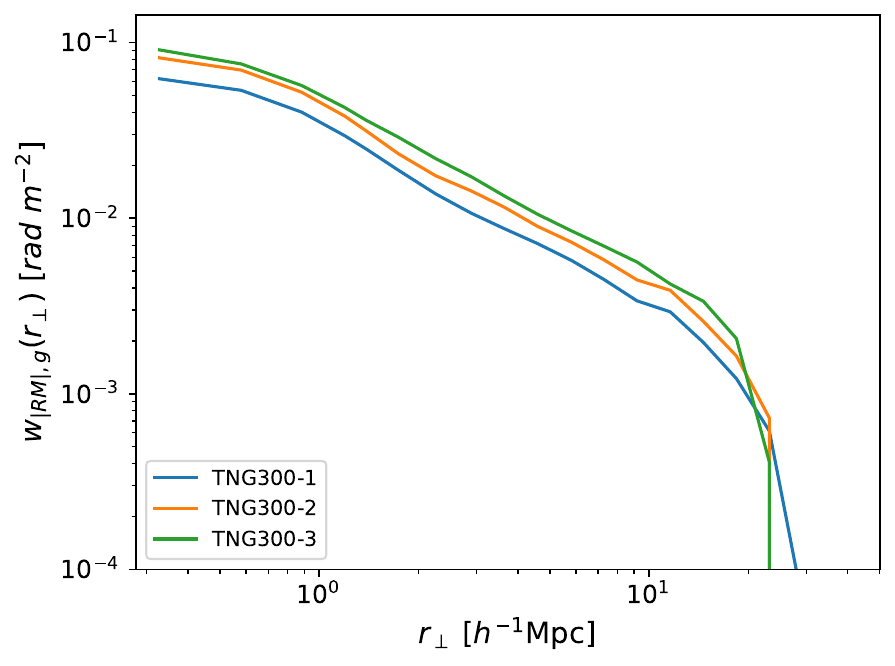}
    \caption{Resolution dependence of the $|\mathrm{RM}|$–halo cross-correlation at $z=0$ for three IllustrisTNG runs: TNG300-1, TNG300-2, TNG300-3. Consistent with the RM$^2$–based statistic, the $|\mathrm{RM}|$–halo correlation function amplitude decreases with increasing resolution.
}
    \label{fig:rm_abs_res}
\end{figure}

\end{appendix}

To explore the impact of simulation resolution, we repeated our analysis using higher-resolution runs among the TNG300 series, TNG300-1 and TNG300-2, which have 64 and 8 times higher resolution than TNG300-3, respectively. Figure~\ref{fig:rm_sq_res} shows the RM$^2$–halo cross-correlation function and its analytical approximation at a representative redshift of $z = 0$. The amplitude of the RM$^2$–halo cross-correlation function decreases with increasing resolution. A similar trend is seen in Figure~\ref{fig:rm_abs_res}, which shows the $|\mathrm{RM}|$–halo cross-correlation function. We attribute this resolution dependence to the direction of magnetic fields in the simulations. In higher resolution runs, the magnetic field direction is better resolved and can vary significantly between neighboring resolution elements, leading to cancellations in the projected electron-density–weighted magnetic field. In contrast, in lower-resolution runs, these small-scale directional fluctuations are not fully captured, resulting in a more coherent field orientation and an artificially enhanced net projection. This effect increases the magnitude of both the |RM| and RM$^2$ fields in low-resolution data. 

This trend is fairly well captured by the analytic approximation (Eq.~\ref{eq:wrm_approx}), although the small-scale agreement is less good at high resolution. Note that, in each case, the analytic calculations adopt the $P_{\tilde{B}}(q_\parallel=0,q_\perp)$ and $P_{e,g}(k)$ power spectra measurements from the simulations at the appropriate resolution. We find that the electron-galaxy cross-power spectrum is approximately unchanged with resolution. Instead, the dominant effect is a suppression of the projected, electron-density--weighted magnetic field strength with increasing resolution.  
The calibration constant is fixed to the low resolution value of $K=3$ -- adjusting this value might impact the precise results here. The results in this section confirm that our main conclusions are robust to simulation resolution. However, we encourage further investigation regarding the sensitivity of these statistics to simulation model, resolution, and boxsize, especially before performing detailed comparisons with observational data. 

\bigskip

\acknowledgments

We thank the anonymous referee and editor for helpful comments, which helped clarify several important points and sharpen our presentation.
We acknowledge the use of publicly available Python package: \texttt{Astropy} \citep{astropy_collaboration_astropy_2013, astropy_collaboration_astropy_2018, astropy_collaboration_astropy_2022},  \texttt{NumPy} \citep{harris_array_2020}, \texttt{SciPy} \citep{virtanen_scipy_2020}, \texttt{Matplotlib} \citep{hunter_matplotlib_2007}, \texttt{Corrfunc} \citep{sinha_corrfunc_2020}, \texttt{Colossus} \citep{diemer_colossus_2018}, and \texttt{Nbodykit}  \citep{hand_nbodykit_2018}



\newpage
\bibliographystyle{JHEP}
\bibliography{main.bib}

\end{document}